%% file: wiretap1.tex
\documentclass[draftcls,onecolumn]{IEEEtran}
\usepackage{amsmath,amsfonts}
\usepackage{color,graphics}
\newtheorem{thm}{Theorem}

\newtheorem{example}{Example}

\begin{document}
\title{Applications of LDPC Codes to the Wiretap Channel} 
\author{Andrew~Thangaraj,~\IEEEmembership{Member,~IEEE,} 
                   Souvik~Dihidar,~\IEEEmembership{Student Member,~IEEE,} 
                   A.~R.~Calderbank,~\IEEEmembership{Fellow,~IEEE,} 
                   Steven~McLaughlin,~\IEEEmembership{Senior~Member,~IEEE,}
                   Jean-Marc~Merolla,~\IEEEmembership{Member,~IEEE}%
\thanks{A. Thangaraj is with the Indian Institute of Technology, Madras}\thanks{A. R. Calderbank is with the Department of Electrical Engineering, Princeton University}\thanks{S. Dihidar, S. McLaughlin, and J.-M. Merolla are with the GTL-CNRS Telecom lab, Metz, France}}
\maketitle
\begin{abstract}
With the advent of quantum key distribution (QKD) systems, perfect (i.e. information-theoretic) security can now be achieved for distribution of a cryptographic key. QKD systems and similar protocols use classical error-correcting codes for both error correction (for the honest parties to correct errors) and privacy amplification (to make an eavesdropper fully ignorant). From a coding perspective, a good model that corresponds to such a setting is the wire tap channel introduced by Wyner in 1975. In this paper, we study fundamental limits and coding methods for wire tap channels. We provide an alternative view of the proof for secrecy capacity of wire tap channels and show how capacity achieving codes can be used to achieve the secrecy capacity for any wiretap channel. We also consider binary erasure channel and binary symmetric channel special cases for the wiretap channel and propose specific practical codes. In some cases our designs achieve the secrecy capacity and in others the codes provide security at rates below secrecy capacity. For the special case of a noiseless main channel and binary erasure channel, we consider encoder and decoder design for codes achieving secrecy on the wiretap channel; we show that it is possible to construct linear-time decodable secrecy codes based on LDPC codes that achieve secrecy. 
\end{abstract}
\IEEEpeerreviewmaketitle
\section{Introduction and Motivation}
The notion of communication with perfect security was defined in information-theoretic terms by Shannon \cite{Shannonsec}. Suppose a $k$-bit message $\mathbf{M}$ is to be transmitted securely from Alice to Bob across a public channel. Perfect security is said to be achieved if the encoding of $\mathbf{M}$ into a transmitted word $\mathbf{X}$ is such that the mutual information $I(\mathbf{M};\mathbf{X})=0$. From this definition, Shannon concluded that Alice and Bob should necessarily share $k$ bits of key for achieving perfect security.

\begin{figure}[ht]
\begin{center}
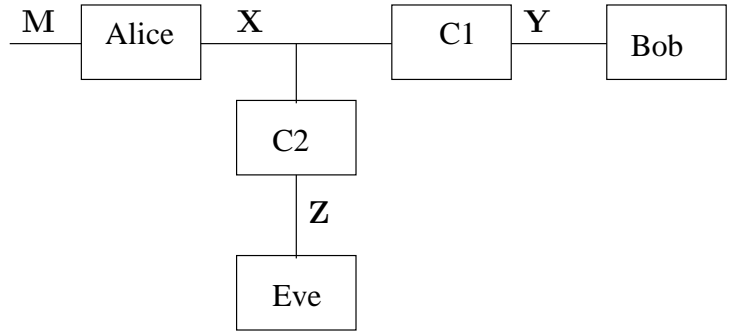
\end{center}
\caption{Wire tap channel.}
\label{fig:BCS}
\end{figure}

An alternative notion of communication with security was introduced by Wyner \cite{Wyner1}. Wyner introduced the wire tap channel, which has matured into a system depicted in Fig. \ref{fig:BCS}. In a wire tap channel, the honest parties Alice and Bob are separated by a channel C1 called the main channel. The important modification when compared to Shannon's study of security is that any eavesdropper Eve observes information transmitted by Alice through another channel C2 called the wiretapper's channel. C1 and C2 are assumed to be discrete memoryless channels (DMCs).
Suppose Alice and Bob try to (securely) communicate a $k$-bit message $\mathbf{M}$  across C1. Alice encodes $M$ into an $n$-bit transmitted word $\mathbf{X}$. The legitimate receiver Bob and an eavesdropper Eve receive $\mathbf{X}$ through two different channels C1 and C2, respectively. Bob's and Eve's observations are denoted $\mathbf{Y}$ and $\mathbf{Z}$, respectively. Alice's encoding should achieve two objectives: (1) [Security] In words, $\mathbf{Z}$ should provide no information about $\mathbf{M}$. The precise formulation used in this paper is that the rate of mutual information $\dfrac{1}{n}I(\mathbf{M};\mathbf{Z})\rightarrow0$ as $n\rightarrow\infty$ (2) [Reliability] $\mathbf{Y}$ can be decoded into $\mathbf{M}$ with negligibly small probability of error. Wyner showed that both objectives can be attained by forward coding without any key bits if the channels C1 and C2 satisfy some conditions. The rate $k/n$ is called the secrecy rate. 

Secrecy capacity of a wire tap channel is the largest $k/n$ for which the objectives of secure and reliable communication is achievable. Secrecy capacity is a function of the channels C1 and C2. If the capacity of C1 is greater than the capacity of C2, one would intuitively expect secrecy capacity to be positive. This intuition has been justified in several cases. Wyner \cite{Wyner1} showed that if C2 is a degraded version of C1 (C2 is C1 concatenated with another DMC) then secrecy capacity is positive. Csisz\'{a}r and K\"{o}rner \cite{Csis} showed that the secrecy capacity is positive for the cases when C1 is ``less noisy'' than C2. However, computing secrecy capacity of a general wire tap channel efficiently given DMCs C1 and C2 still remains an unsolved problem. The most recent progress in this problem was made by Van Dijk \cite{Dijk}.

The key distribution problem in wire tap channels, which falls under the general problem of key generation from correlated source outputs, has been studied extensively \cite{maurer93,maurer99,ahlswede}. The objective of secure key distribution is for Alice and Bob to share a common $k$-bit key about which Eve's entropy is maximal. In key distribution, the $k$ bits can be unknown to Alice before transmission. Powerful ideas such as common randomness, advantage distillation and privacy amplification were developed in the context of key distribution over wire tap channels \cite{ahlswede,bennett95:IEEE_tit}. Several key distribution protocols have been developed and studied; many of the protocols make use of a parallel, error-free public channel between Alice and Bob during implementation.

The problem of developing forward coding schemes (with no parallel channel) for secure communication over wire tap channels has not received much attention. Some examples of coding schemes have been provided in \cite{Wyner1} and \cite{maurer93}. A condition for constructing codes for the modified wire tap channel, introduced by Ozarow and Wyner \cite{Wyner2}, has been studied by Wei \cite{Wei}. Code construction methods and their connection to security have not been extensively explored so far. However, existence of coding schemes for various generalized wire tap channel scenarios has been proved by several authors recently \cite{hayashi, hayashipp, muramatsu}. In particular, the existence of coding methods based on LDPC codes has been shown in \cite{muramatsu}.

In this paper, we focus on the problem of developing coding schemes for secure communication across wire tap channels. We begin by discussing the secrecy capacity theorem for certain wire tap channels. We provide a careful reworking of the proof so that the requirements of security and reliability are separated. We generalize an important link between capacity-approaching codes and security. This alternative view of the proof provides a clear construction method for coding schemes for secure communication across arbitrary wire tap channels.

Later, we use this idea to develop codes for different wire tap channels. For a wire tap channel with a noiseless main channel and a binary erasure channel (BEC) as the wire tapper's channel, we provide codes that achieve secrecy capacity using the threshold properties of codes on graphs under message passing decoding. To our knowledge, these are the first codes that achieve secrecy capacity over wire tap channels. Using this construction, we show that it is possible to construct linear-time decodable codes that achieve security over such wire tap channels. Next, we extend the construction to wire tap channels that have BECs as both the main and wiretapper's channel. We show important connections between the threshold of codes on graphs under message-passing decoding and security. Finally, we consider a wire tap channel with a noiseless main channel and a binary symmetric channel (BSC) as the wiretapper's channel. For this case, we provide a coding solution using codes that have good error-detecting capability. 

Throughout the paper, the criterion for security is that the mutual information between the message and an eavesdropper's observables goes to zero {\it rate-wise}. Note that this formulation (originally due to Wyner \cite{Wyner1}) is weaker than the accepted security criteria in contemporary work in cryptography, which typically require the total mutual information to go to zero. Hence, this work can be seen as a conceptual advancement in the area of forward coding for wire tap channels. With future study, stronger security criterion such as exponential fall in mutual information could become possible for such codes over a wire tap channel.

The rest of the paper is organized as follows: In Section \ref{sec:WTC}, we briefly discuss secrecy capacity for wire tap channels and point out the connection between capacity-approaching codes and secrecy. In Section \ref{sec:BEC}, we discuss the general coding scheme for wire tap channels used in the remainder of the paper. In Section \ref{sec:codmetewt}, we present codes for wire tap channels with a noiseless main channel and a BEC as the wiretapper's channel; in Section \ref{sec:efficientdecode}, we modify the above codes and extend them to construct linear-time decodable codes for these wire tap channels. In Section \ref{sec:codmet2}, we present codes for wire tap channels with BECs as both main and wiretapper's channels. In Section \ref{sec:BSC}, we present code constructions for wire tap channels with a noiseless main channel and a BSC as the wiretapper's channel. Finally, we conclude in Section \ref{sec:conclusion} with a discussion of results and topics for future research.
\section{Coding for the Wire Tap channel}
\label{sec:WTC}
In a general wire tap channel (Fig. \ref{fig:BCS}), C1 and C2 are discrete memoryless channels (DMCs). The two DMCs have the same input alphabet but different output alphabet. C1 is denoted $X\rightarrow Y$, where $X$ is a random variable denoting an input symbol to C1, and $Y$ is a random variable denoting an output symbol from C1. Similarly, C2 is denoted $X\rightarrow Z$. A sequence of $n$ input symbols is denoted by $X^n$ or $\mathbf{X}$. $Y^n$ and $\mathbf{Y}$, and $Z^n$ and $\mathbf{Z}$ have similar notations for the outputs. C1 and C2 of a wire tap channel are called the main channel and wire tap channel, respectively.
\subsection{Secrecy capacity of the wire tap channel}
The notion of secrecy capacity, as introduced by Wyner \cite{Wyner1}, has an operational meaning of being the maximum possible rate of information transmission between Alice and Bob that still enables Eve to be kept totally ignorant. Before defining the operational meaning precisely, we look at the calculation of secrecy capacity for a given wire tap channel. The secrecy capacity $\text{C}_s$ for a general wire tap channel can be calculated as follows \cite{Csis}:
\begin{equation}
\text{C}_s=\max_{V\rightarrow X\rightarrow(Y,Z)}\left[I(V;Y)-I(V;Z)\right],
\label{eqn:Cs}
\end{equation}
where the maximum is over all possible random variables $V$ in joint distribution with $X$, $Y$ and $Z$ such that $V\rightarrow X\rightarrow(Y,Z)$ is a Markov chain. The random variable $V$ does not have a direct physical meaning; it is used for calculation purposes. Note that $\text{C}_s$ could turn out to be zero or negative in some cases. At present, the calculation of secrecy capacity is an unsolved problem when C1 and C2 are general DMCs. However, the calculation of secrecy capacity can be simplified for some special cases that impose restrictions on the wire tap channel with respect to the main channel.

If $I(V;Y)\ge I(V;Z)$ for all Markov chains $V\rightarrow X\rightarrow(Y,Z)$, the main channel is said to be less noisy than the wire tap channel. If the main channel is less noisy than the wire tap channel \cite{Csis}, then
\begin{equation}
\text{C}_s=\max_{P_{X}(x)}\left[I(X;Y)-I(X;Z)\right],
\label{eqn:Csln}
\end{equation}
where the maximum is over all possible distributions $P_{X}(x)$ of $X$. Moreover, as shown in \cite{Dijk}, $I(X;Y)-I(X;Z)$ is a convex function of $P_{X}(x)$ when the main channel is less noisy than the wire tap channel; hence, the secrecy capacity can be calculated using convex optimization methods. It was further shown in \cite{Dijk} that if $I(X;Y)$ and $I(X;Z)$ are individually maximized by the same $P_{X}(x)$, and the main channel ($X\rightarrow Y$) is less noisy than the wire tap channel ($X\rightarrow Z$), then
\begin{equation}
\text{C}_s=\text{Capacity}(X\rightarrow Y)-\text{Capacity}(X\rightarrow Z),
\label{eqn:Cssym}
\end{equation}
where Capacity(.) refers to the usual channel capacity.
\subsection{Coding method}
\label{sec:codmet}
The coding problem for Alice in the wire tap channel involves adding redundancy for enabling Bob to correct errors (across the main channel) and adding randomness for keeping Eve ignorant (across the wire tap channel). The coding method presented here is not new. It is present in the proofs in \cite{Wyner1} and \cite{Csis}. More recently, similar coding methods have been used in \cite{hayashi, hayashipp} for finding bounds and error exponents in the context of wire tap channels. However, our method of proof separates the requirements of security and reliability and results in a simple design method for codes over a wire tap channel.

\begin{figure}[htb]
\begin{center}
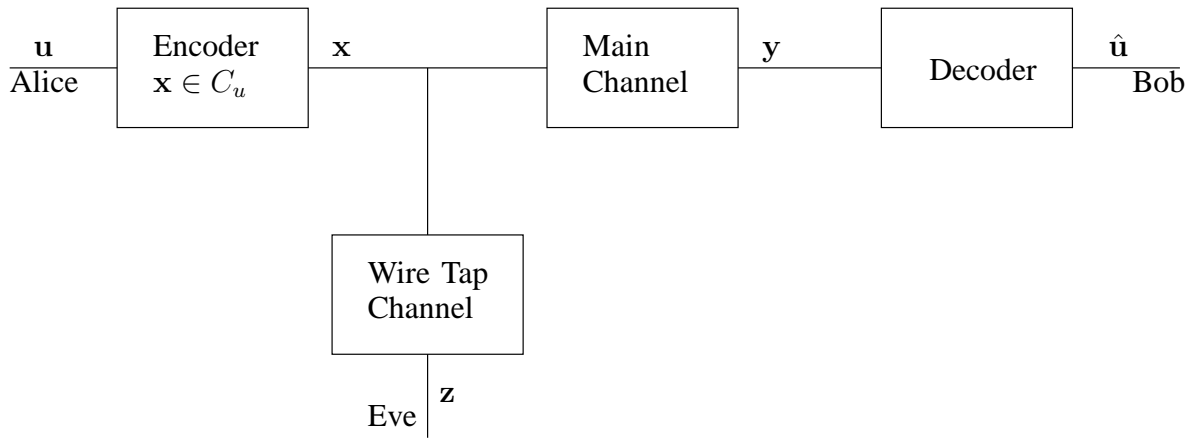
\end{center}
\caption{Coding method for the wire tap channel.}
\label{fig:codmet}
\end{figure}

Let us assume that Alice needs to transmit one out of $M$ equally likely messages i.e. a message denoted $u$ is such that $u\in\{1,2,\cdots,M\}$ and Prob$\{u=i\}=1/M$. Alice uses $M$ codes $C_i,\;1\leq i\leq M$ with $|C_i|=L$ and block-length $n$. Each codeword of $C_i$ consists of $n$ symbols from the input alphabet of the main or wire tap channel. We let the common input alphabet to the two channels be $\{1,2,\cdots,K\}$. A symbol of the input alphabet is denoted $k$. A message $u$ is encoded into a transmitted word $\mathbf{x}$ as follows: $\mathbf{x}$ is chosen uniformly at random from the code $C_u$. The coding method is illustrated in Fig. \ref{fig:codmet}. The transmitted word $\mathbf{x}$ , in general, belongs to the overall code $C=\cup_i C_i$. The rate of information transmission from Alice to Bob (in terms of bits per channel use) in such a setting is given by $\log_2M/n$. The receiver on the main channel (Bob) decodes a received word $\mathbf{y}$ with respect to the overall code $C$ into a decoded message $\hat{u}$ (say, by Maximum-Likelihood (MaxL) decoding). 

The objective of Alice and Bob in a wire tap channel can now be given a precise definition. Let $\mathbf{U}$, $\mathbf{\hat{U}}$, and $\mathbf{Z}$ be random variables denoting Alice's message, Bob's decoded message, and Eve's received word, respectively. Let $H(V)$ represent the entropy of a random variable $V$. Then, the objective is to achieve the following:
\begin{eqnarray}
\label{eqn:Pe}\text{Prob}\{\mathbf{U}\ne\mathbf{\hat{U}}\}&\rightarrow&0.\\
\label{eqn:H}I(\mathbf{U};\mathbf{Z})/n&\rightarrow&0.
\end{eqnarray}
The constraint (\ref{eqn:H}) is referred to as the security constraint, while (\ref{eqn:Pe}) is called the reliability constraint. If an encoder (as in Fig. \ref{fig:codmet}) with $R_s=\log_2M/n$ satisfies the security and reliability constraints for a given wire tap channel, then such an encoder is said to achieve a secrecy rate $R_s$.
\subsection{Security of the coding method}
\label{sec:seccon}
The security constraint is of paramount importance in the design of an encoder for a wire tap channel. The following choice of the codes $C_{u}$ satisfies the security constraint: Each $C_{u}$ should approach capacity over the wire tapper's channel (similar to the special case considered by Wyner in \cite{Wyner1}). We present the criterion in the following theorem (the notation used is from Fig. \ref{fig:codmet} and Section \ref{sec:codmet}).
\begin{thm}
If each code $C_u,u\in\{1,2,\cdots,M\}$ comes from a sequence of codes that approach capacity asymptotically over the wire tap channel, then $I(\mathbf{U};\mathbf{Z})/n\rightarrow0$, as $n\rightarrow\infty$.
\label{thm:seccon}
\end{thm}
\begin{proof}
Since each $C_u$ approaches the capacity $C_w$ of the wire tapper's channel, we have for any $\epsilon>0$ an $n_{\epsilon}$ such that for $n>n_{\epsilon}$, $I(\mathbf{X};\mathbf{Z}|\mathbf{U}=u)/n\ge C_W-\epsilon$ for each $u$. Therefore for $n>n_{\epsilon}$, $I(\mathbf{X};\mathbf{Z}|\mathbf{U})/n\ge C_W-\epsilon$.

Expanding $I(\mathbf{Z};\mathbf{U}\mathbf{X})$ in two ways, we get
$$I(\mathbf{Z};\mathbf{U}\mathbf{X})=I(\mathbf{U};\mathbf{Z})+I(\mathbf{X};\mathbf{Z}|\mathbf{U})=I(\mathbf{X};\mathbf{Z})+I(\mathbf{U};\mathbf{Z}|\mathbf{X}).$$
Since $\mathbf{U}\rightarrow \mathbf{X}\rightarrow \mathbf{Z}$ is a Markov chain, $I(\mathbf{U};\mathbf{Z}|\mathbf{X})=0$. Therefore for $n>n_{\epsilon}$ we have
$$I(\mathbf{U};\mathbf{Z})/n=I(\mathbf{X};\mathbf{Z})/n-I(\mathbf{X};\mathbf{Z}|\mathbf{U})/n\le C_W-(C_W-\epsilon)=\epsilon.$$
\end{proof}
This fundamental connection between capacity-approaching codes and secrecy has been used in many works on wire tap channels beginning with \cite{Wyner1} implicitly. In Appendix \ref{app:proof2}, we show that this connection can be used to design codes that approach the secrecy capacity of certain wire tap channels. Particularly, we have shown that the reliability condition can be satisfied while simultaneously forcing each code $C_u$ to approach capacity. 

In summary, we have shown that secrecy capacity can be achieved for certain wire tap channels using codes that achieve capacity over the wire tapper's channel. A significant drawback is that capacity-achieving codes are essential for guaranteeing the security of the method. Since capacity-achieving codes are not practical in many settings, design of practical codes that are secure is an important problem that needs to be addressed. If the resulting code is practical and secure, transmission rates below secrecy capacity are certainly acceptable. The remainder of this paper is concerned with developing practical codes and protocols for wire tap channels. In some simple settings, practical methods that achieve secrecy capacity are given.
\section{Code Design for the Wire Tap channel}
\label{sec:BEC}
In this section, we study the design and use of linear codes over a wire tap channel. We use a method that was first introduced and studied by Wyner and Ozarow \cite{Wyner1, Wyner2} for two specific cases. We have extended Wyner's study by considering other wire tap channels. We have also provided better, implementable codes for the cases studied by Wyner.
\subsection{Coding method}
\label{sec:BECcodmet}
We consider a coding method similar to Fig. \ref{fig:codmet} but with linear codes and cosets. To transmit $k$-bit messages, we first select a $(n,l)$ linear binary code $C$ such that $k\le n-l$. Out of the $2^{n-l}$ cosets of $C$, we choose $2^k$ cosets and let each message correspond to a chosen coset. The selection of the cosets is done in a linear fashion. Suppose $G$ is a generator matrix for $C$ with rows $\mathbf{g}_1$, $\mathbf{g}_2$, $\cdots$, and $\mathbf{g}_l$. We select $k$ linearly independent vectors $\mathbf{h}_1$, $\mathbf{h}_2$, $\cdots$, and $\mathbf{h}_k$ from $\{0,1\}^n\setminus C$. The coset corresponding to a $k$-bit message $\mathbf{s}=[s_1\;s_2\cdots s_k]$ is determined as follows:
\begin{equation}
\mathbf{s}\rightarrow s_1\mathbf{h}_1+s_2\mathbf{h}_2+\cdots+s_k\mathbf{h}_k+C.
\label{eqn:mtoc}
\end{equation}
Though the above correspondence is deterministic, the encoding procedure has a random component in the selection of the transmitted word. A $k$-bit message $\mathbf{s}$ is encoded into a $n$-bit word randomly selected from the coset of $C$ corresponding to $\mathbf{s}$. Hence, the transmitted word, $\mathbf{x}$, is given by 
$$\mathbf{x}=s_1\mathbf{h}_1+s_2\mathbf{h}_2+\cdots+s_k\mathbf{h}_k+v_1\mathbf{g}_1+v_2\mathbf{g}_2+\cdots+v_l\mathbf{g}_l,$$ 
where $\mathbf{v}=[v_1\;v_2\cdots v_l]$ is an uniformly random $l$-bit vector. The overall encoding operation can be described as a matrix multiplication. Let $\mathbf{G^*}$ be the $k\times n$ matrix with rows $\mathbf{h}_1$, $\mathbf{h}_2$, $\cdots$, and $\mathbf{h}_k$. Then,
$$\mathbf{x}=[\mathbf{s}\;\mathbf{v}]\begin{bmatrix}G^*\\G\end{bmatrix}.$$
Hence, $\mathbf{x}$ belongs to the code $\overline{C}$ with generator matrix
$$\overline{G}=\begin{bmatrix}G^*\\G\end{bmatrix}.$$
The goal of both the legitimate receiver and the eavesdropper is to determine $\mathbf{s}$ from their respective received vectors. Restating the conditions of Section \ref{sec:codmet}, the design of the codes $C$ and $\overline{C}$ should be such that (1) $\mathbf{s}$ can be determined without error across the main channel, and (2) every $\mathbf{s}$ is equally likely across the wiretapper's channel.

Guided by the results of the previous sections, we could choose $C$ as a capacity-achieving code over the wiretapper's channel. However, designing a code $\overline{C}$ that can be decoded across the main channel is still a challenge. Moreover, capacity-achieving codes have not yet been demonstrated in practice for many channels. In the following sections, we look at some design approaches for some simple wire tap channels. The encoding method and notation will remain the same for all cases.
\section{Noiseless main channel and Erasure wiretapper's channel}
\label{sec:codmetewt}
\begin{figure}[htb]
\begin{center}
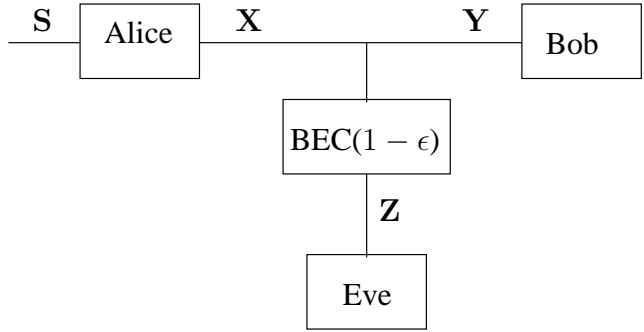
\end{center}
\caption{Wire tap channel denoted EWT$(\epsilon)$.}
\label{fig:ewt}
\end{figure}

We begin with the simplest possible wire tap channel with a binary erasure channel (BEC) as the wiretapper's channel and a noiseless main channel. This scenario is shown in Fig. \ref{fig:ewt}.
In Fig. \ref{fig:ewt}, the wiretapper's channel has been denoted BEC$(1-\epsilon)$ i.e. the probability of erasure in the wiretapper's channel is $1-\epsilon$. The probability that a bit is leaked to the wiretapper is $\epsilon$. This notation has been chosen for future convenience. We will denote the wire tap channel of Fig. \ref{fig:ewt} as EWT$(\epsilon)$. Using (\ref{eqn:Cssym}), we see that the secrecy capacity of an EWT$(\epsilon)$ is
\begin{equation}
\text{C}_s=1-\text{Capacity}(\text{BEC}(1-\epsilon))=1-(1-(1-\epsilon))=1-\epsilon.
\label{eqn:CsBEC}
\end{equation}

\begin{figure}[htb]
\begin{center}
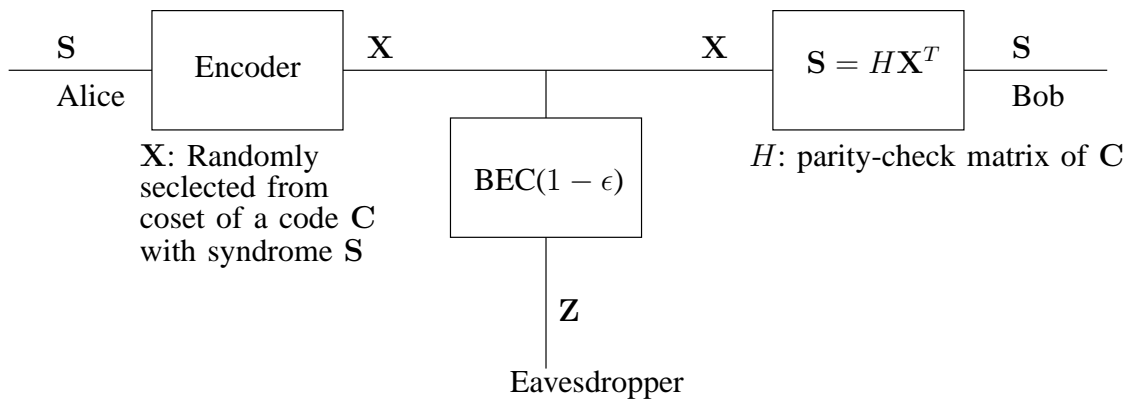
\end{center}
\caption{Coding method.}
\label{fig:wtc2}
\end{figure}

The coding method across an EWT$(\epsilon)$ is illustrated in Fig. \ref{fig:wtc2}.
In the figure, $\mathbf{S}$ is the random variable denoting the $k$-bit message to be transmitted. The code $C$ is chosen to be an $(n,n-k)$ code, and the code $\overline{C}$ is chosen to be the entire vector space $\{0,1\}^n$. The transmitted $n$-tuple is denoted by the random variable $\mathbf{X}=[X_1\;X_2\cdots X_n]$. Note that the message $\mathbf{S}$ can be seen as a syndrome of $C$ with respect to a carefully constructed $k\times n$ parity-check matrix ${H}$. Since the channel between Alice and Bob is error-free, Bob finds the message as follows: $\mathbf{S}={H}\mathbf{X}^T\pmod{2}$. The secret information rate is $R=k/n$. From (\ref{eqn:CsBEC}), we see that for secure transmission,
\begin{equation}
R=k/n<1-\epsilon.
\label{eqn:ubar}
\end{equation}
Assuming that all messages are equally likely, we have $X_i=0$ or $X_i=1$ with probability $1/2$ each. The eavesdropper learns $X_i$ with probability $\epsilon$. That is, the random variable $\mathbf{Z}=[Z_1\;Z_2\cdots Z_n]$ is such that $Z_i=X_i$ with probability $\epsilon$, and $Z_i=?$ (unknown or erasure) with probability $1-\epsilon$. 
\subsection{Security Criterion}
\label{sec:security}
To develop a security criterion for the choice of $C$, we calculate the eavesdropper's uncertainty $H(\mathbf{S}|\mathbf{Z})$ by first evaluating $H(\mathbf{S}|\mathbf{Z}=\mathbf{z})$. Note that the eavesdropper is given complete knowledge of the code $C$ and infinite computational power. The main source of uncertainty is the uniformly random selection of the transmitted word $\mathbf{X}$ from the coset of $C$ corresponding to the message $\mathbf{S}$.

If a coset of $C$ contains at least one vector that agrees with $\mathbf{z}\in\{0,1,?\}^n$ in the unerased positions, we say that the coset is consistent with $\mathbf{z}$. Each consistent coset corresponds to a possible message for the eavesdropper. Let $\mathbf{v}$ be a vector consistent with $\mathbf{z}$ in the coset $\mathbf{v}+C$. Let $S$ be the set of all vectors in $\mathbf{v}+C$ consistent with $\mathbf{z}$. Then, $\mathbf{v}+S$ is the set of all vectors in $C$ with zeros in the positions revealed in $\mathbf{z}$. That is,
$$\mathbf{v}+S=\{\mathbf{u}\in C: u_i=0\text{ whenever }z_i\ne?\}.$$
Since $|S|=|\mathbf{v}+S|$, the number of vectors consistent with $\mathbf{z}$ in each consistent coset is a constant equal to the size of the set on the RHS above.

Let $N(C,\mathbf{z})$ denote the total number of cosets of $C$ consistent with $\mathbf{z}$. Since each message is equally likely {\it{a priori}}, we get
\begin{equation}
\label{eqn:HSz}
H(\mathbf{S}|\mathbf{Z}=\mathbf{z})=\log_2 N(C,\mathbf{z}).
\end{equation}

For an $(n,n-k)$ code $C$, the maximum possible value for $N(C,\mathbf{z})$ is the total number of cosets $2^k$. If $N(C,\mathbf{z})=2^k$, we say that $\mathbf{z}$ is secured by $C$ since the eavesdropper's Prob$\{\mathbf{S}=\mathbf{s}|\mathbf{Z}=\mathbf{z}\}=1/2^k$ for every possible message $\mathbf{s}$. The following theorem (adapted from \cite[Lemma 3]{Wyner2}) states a condition for a vector $\mathbf{z}$ to be secured by a code $C$.
\begin{thm}[Ozarow, Wyner '84]
\label{thm:1}
Let an $(n,n-k)$ code $C$ have a generator matrix $G=[\mathbf{a}_1\cdots \mathbf{a}_n]$, where $\mathbf{a}_i$ is the $i$-th column of $G$. Consider an instance of the eavesdropper's observation $\mathbf{z}\in\{0,1,?\}^n$ with $\mu$ unerased positions given by $\{i:\mathbf{z}_i\ne ?\}=\{i_1,i_2,\cdots,i_{\mu}\}$. $\mathbf{z}$ is secured by $C$ iff the matrix $G_{\mu}=[\mathbf{a}_{i_1}\mathbf{a}_{i_2}\cdots \mathbf{a}_{i_{\mu}}]$ has rank $\mu$.
\end{thm}
\begin{proof}
If $G_{\mu}$ has rank $\mu$, the code $C$ has all $2^{\mu}$ possible $\mu$-tuples in the $\mu$ unerased positions. So each coset of $C$ also has all $2^{\mu}$ possible $\mu$-tuples in the $\mu$ revealed positions. So $N(C,\mathbf{z})=2^k$.

If $G_{\mu}$ has rank less than $\mu$, the code $C$ does not have all $\mu$-tuples in the $\mu$ unerased positions. So there exists at least one coset that does not contain a given $\mu$-tuple in the $\mu$ unerased positions, and $N(C,\mathbf{z})<2^k$.
\end{proof}
If a random vector obtained over a BEC$(1-\epsilon)$ is secured with probability close to one by an $(n,n-k)$ code $C$, rate $k/n$ is achievable with secrecy over an EWT$(\epsilon)$.
\subsection{Using duals of codes on graphs}
\label{sec:ldpc}
We now study the use of the threshold property of codes on graphs for providing security over an erasure wire tap channel. We illustrate the method using Low-Density Parity-Check (LDPC) codes. The extension to other codes on graphs is shown in examples.

Consider a bipartite graph ensemble $C^n(\lambda,\rho)$ with $n$ left nodes and left and right edge degree distribution polynomials $\lambda(x)=\sum_{i\geq 1}\lambda_{i}x^{i-1}$ and $\rho(x)=\sum_{i\geq 1}\rho_{i}x^{i-1}$, respectively \cite{effenc}. The coefficients $\lambda_i$ (respectively, $\rho_i$) denotes the probability that a randomly chosen edge in the Tanner graph of the code is incident on a variable (respectively, check) node of degree $i$. The adjacency matrix of a graph from the ensemble provides the parity-check matrix of a low-density parity-check (LDPC) code. Let the threshold for $C^n(\lambda,\rho)$ over the binary erasure channel be $\alpha^*(\lambda,\rho)$. The threshold property has the following straight-forward interpretation:
\begin{thm}
\label{thm:ldpc}
Let ${M}$ be a parity-check matrix of an LDPC code from the ensemble $C^n(\lambda,\rho)$. A submatrix formed by selecting columns of ${M}$ independently with probability $\alpha$ will have full column rank for $\alpha<\alpha^*(\lambda,\rho)$ for large $k$ with high probability.
\end{thm}
Theorem \ref{thm:ldpc} enables the use of duals of LDPC codes as the code $C$ over an EWT($\epsilon$) as shown in Fig. \ref{fig:wtc2}. We let a matrix ${M}$ from the ensemble $C^n(\lambda,\rho)$ to be the generator matrix for $C$. By Theorem \ref{thm:ldpc}, the columns of the matrix ${M}$ corresponding to the leaked bits over a BEC$(1-\epsilon)$ will have full rank with high probability whenever $\epsilon<\alpha^*(\lambda,\rho)$. Note that the probability that a bit is leaked across a BEC$(1-\epsilon)$ is equal to $\epsilon$. In combination with Theorem \ref{thm:1}, we see that the code $C$ with generator matrix ${M}$ provides complete security with probability tending to one for large block-length over an EWT$(\epsilon)$ with $\epsilon<\alpha^*(\lambda,\rho)$.
\begin{example}
\label{ex:ldpc}
The $C^n(x^2,x^5)$ ensemble of $(3,6)$-regular LDPC codes has threshold $\alpha^*(x^2,x^5)\approx0.42$. Let $M$ be an adjacency matrix from the ensemble with large $n$ (say, $n>10^5$). $M$ is an $n/2\times n$ binary matrix with row weight $3$ and column weight $6$. The $(n,n/2)$ code $C$ with generator matrix $M$ can be used over an EWT$(\epsilon)$ for $\epsilon<0.42$ with secrecy. The information rate between the honest parties in this case is $R=0.5$ compared to the upper bound of $1-\epsilon=0.58$ (from (\ref{eqn:ubar})). (In practice, the value of $\epsilon$ could be reasonably lesser than $0.42$ for added security.)
\end{example}

The above argument can be extended to other ensembles of codes on graphs that have capacity-achieving thresholds over the binary erasure channel. We illustrate the method with the following example.
\begin{example}[Tornado codes]
\label{ex:tornado}
A rate-$2/3$ tornado code ensemble with threshold $\delta=0.33257$ has been reported in \cite{capsequence}. A parity-check matrix $M$ for a code from the ensemble will have dimensions $n/3\times n$. The $(n,n-2/3n)$ code $C$ with generator matrix $M$ can be used over an EWT$(\epsilon)$ for $\epsilon<0.33257$ with secrecy. The information rate between the honest parties in this case is $R=2/3=0.66666...$ compared to the upper bound of $1-\epsilon=0.66743$.
\end{example}
Similar examples using the other classes of capacity-approaching ensembles can be constructed. Hence over an erasure wire-tap channel with wire-tap probability $\epsilon$, secure information transmission rates tending to the upper bound of $1-\epsilon$ are achievable using duals of codes on graphs that approach capacity over the binary erasure channel.

Note that the code $C$ has properties that are opposite to the requirements of Section \ref{sec:codmet}. While we had proposed to use a code that is capacity-achieving over the wiretapper's channel in Section \ref{sec:codmet}, we have used the dual of a capacity-achieving code when the wiretapper's channel is a BEC. In fact, using the dual appears to be a more powerful method since security does not depend on capacity-achieving codes. This observation agrees with the results of \cite{Wyner2}, and both possibilities are worth exploring in other wire tap channels.
\section{Efficiently Decodable Secrecy codes for noiseless main channel and erasure wiretap channel systems}
\label{sec:efficientdecode}
The main advantage of using LDPC codes for error correction over regular erasure channels is that the decoding algorithm is of linear complexity in blocklength \cite{effenc}. This property can be extended to the use of LDPC codes over the erasure wire tap channels as well. We now discuss designing linear-time decodable secrecy codes for the system shown in Fig. \ref{fig:ewt}, where the main channel is noiseless and the wiretap channel is a BEC. 

In the previous section, we showed how to use dual codes of LDPC codes to construct secrecy codes for this system. The cosets of a dual code of an LDPC code are used to send secret messages. Let $C$ be an LDPC code. Let $G$ be the generator matrix of $C^{\perp}$ (i.e. the parity check matrix of $C$), and $H$ be the parity check matrix of $C^{\perp}$ (i.e. the generator matrix of $C$). $G$ is a sparse matrix since $C$ is an LDPC code. As we discussed in the previous sections, a coset of $C$ is indexed by a secret message $\mathbf{S}$ and the transmitted word $\mathbf{X}$ is a randomly chosen word from that coset. 
\begin{figure}
\begin{center}
\scalebox{0.7}{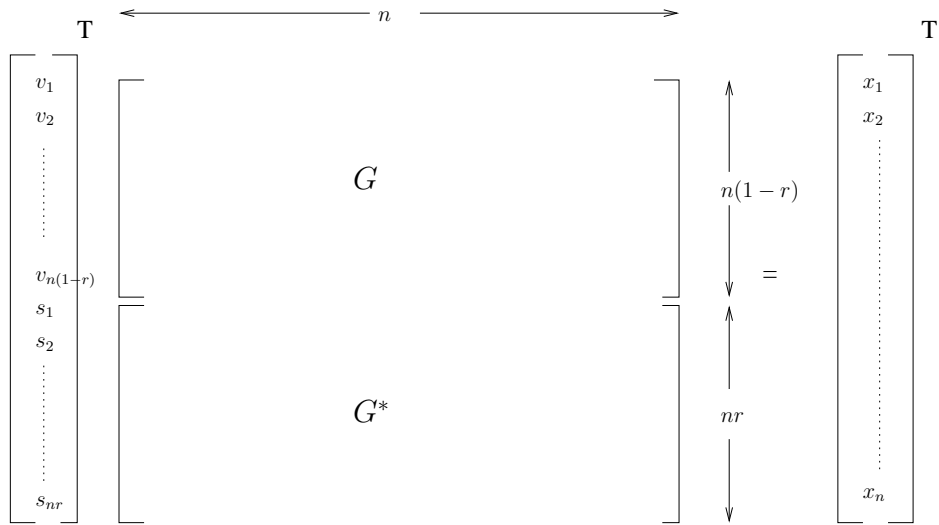}
\end{center}
\caption{The encoding procedure}
\label{fig:contfig}
\end{figure}
Let $C$ have rate $r$, and let $G^{*}$ be the matrix containing the rest of the independent vectors in $\{0,1\}^n$ (as in Section \ref{sec:BECcodmet}). In Fig. \ref{fig:contfig}, we show the matrices $G$ and $G^{*}$, and the method for encoding an $nr$-bit secret message. The bits $\{s_i\}^{nr}_1$ are the secret bits, and the bits $\{v_i\}^{n(1-r)}_1$ are chosen at random. $\{x_i\}^n_1$ are the transmitted bits. We refer to the secret bits, random bits and transmitted bits as $s$-bits, $v$-bits and $x$-bits, respectively.

We now consider the decoding problem for Bob. Suppose $G$ has rows $\mathbf{g}_1$, $\mathbf{g}_2$, $\cdots$, and $\mathbf{g}_{n(1-r)}$. We select $nr$ linearly independent rows $\mathbf{h}_1$, $\mathbf{h}_2$, $\cdots$, and $\mathbf{h}_{nr}$ from $\{0,1\}^n\setminus C$. Let $G^*$ be the matrix with rows as $\mathbf{h}_1$, $\mathbf{h}_2$, $\cdots$, and $\mathbf{h}_{nr}$. Let the matrix $[{G^*}^T,G^T]$ be defined as, 
$$[{G^*}^T,G^T]=[\mathbf{h}^T_1,\mathbf{h}^T_2,\cdots,\mathbf{h}^T_{nr},\mathbf{g}^T_1,\mathbf{g}^T_2,\cdots,\mathbf{g}^T_{n(1-r)}].$$ 
Let $\mathbf{W}$ be defined as $\mathbf{W}=[\mathbf{S},\mathbf{V}]^T$, where $\mathbf{S}$ is the secret message, and $\mathbf{V}$ is a random vector. The transmitted word $\mathbf{X}^T$ is now given by,
\begin{equation}
\mathbf{X}^T=[{G^*}^T,G^T]\mathbf{W}.
\label{eqn:xeqn}
\end{equation}
The decoding problem for Bob is to determine $\mathbf{W}$ (or just $\mathbf{S}$) from (\ref{eqn:xeqn}); this can be easily seen to be a $O(n^2)$ operation. 

We defined $H$ to be a parity check matrix of the code $C^{\perp}$ above. An equivalent way of finding $\mathbf{S}$ is for Bob to compute the syndrome $H\mathbf{X}^T$. For a suitable choice of $H$, one could have $\mathbf{S}=H\mathbf{X}^T$. However, since $H$ is a dense matrix (in general), the complexity of this decoding is also likely to be $O(n^2)$.

We now ask the following question. Is it possible to achieve linear or almost-linear decoding complexity for Bob by carefully choosing a subset of the set of all the cosets of $C$? In other words, can we sacrifice some secrecy rate to achieve linear or almost-linear decoding complexity for Bob? As we will discuss later, our approach will be to make some of the elements of $\mathbf{S}$ always equal to zero. This will decrease our secrecy rate, but we will show that almost-linear or linear time decoding becomes possible in that case.

We first show that the decoding problem in (\ref{eqn:xeqn}) is similar to the problem of systematic encoding of linear block codes. Let a linear block code have parity check matrix $H$, where $H=[H_1,H_2]$. Let the transmitted codeword be $\mathbf{c}=[\mathbf{m}^T,\mathbf{p}^T]^T$, where $\mathbf{m}$ is the message and $\mathbf{p}$ is the parity part. Hence, to find $\mathbf{p}$, the encoder has to solve
\begin{equation}
-H_1\mathbf{m}=H_2\mathbf{p}.
\label{eqn:h1h2}
\end{equation}
Equation (\ref{eqn:xeqn}) is similar to (\ref{eqn:h1h2}) if we let $\mathbf{X}^T=-H_1\mathbf{m}$, $H_2=[{G^*}^T,G^T]$, and $\mathbf{W}=\mathbf{p}$. In \cite{effenc}, the authors have shown how to efficiently solve (\ref{eqn:h1h2}) for LDPC codes. As in \cite{effenc}, our approach will be to multiply (\ref{eqn:xeqn}) by a matrix $Q$ to get
\begin{equation}
Q\mathbf{X}^T=Q[{G^*}^T,G^T]\mathbf{W}.
\end{equation}
To make the operation of finding $\mathbf{W}$ from the above equation $O(n)$, we need to have the matrix $Q[{G^*}^T,G^T]$ in a special form, and we need to ensure that $Q\mathbf{X}^T$ is a $O(n)$ operation.

\begin{figure}[h]
\centering
\scalebox{0.7}{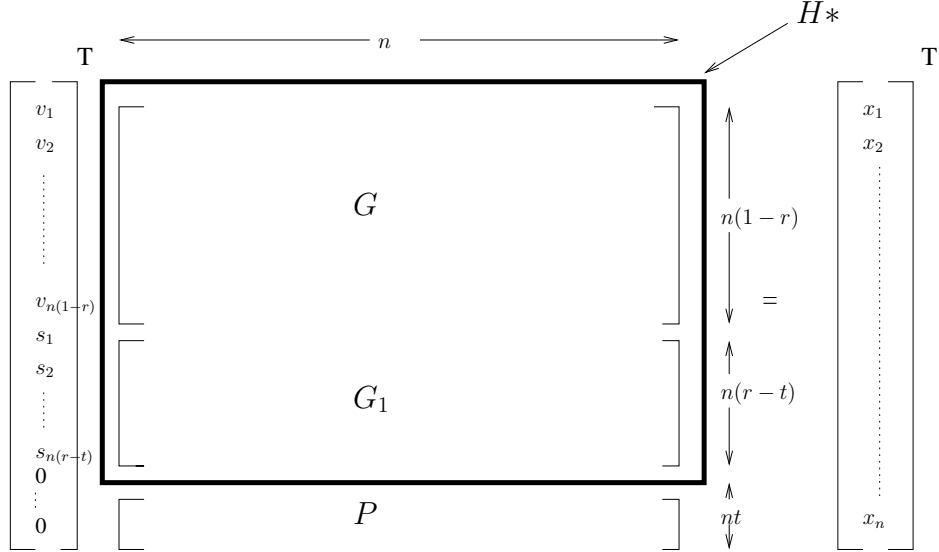}
\caption{Choosing a subset of the set of cosets}
\label{fig:hstar}
\end{figure}

\begin{figure}[h]
\centering
\scalebox{0.7}{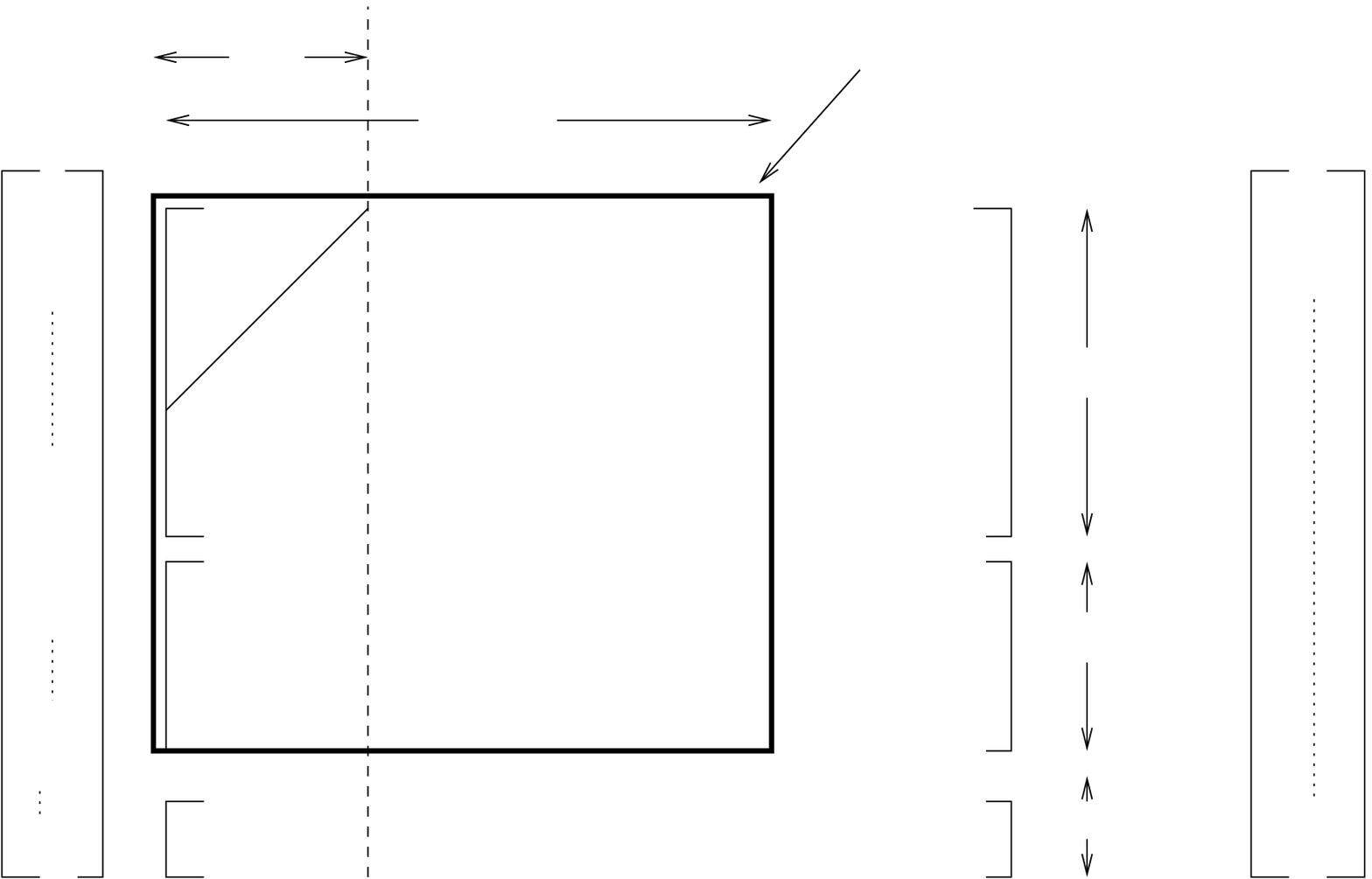}
\caption{The matrix $H_1$}
\label{fig:hbta}
\end{figure}

\subsection{Choosing a subset of the set of cosets}
Let $G_1$ be a sparse, full row-rank matrix whose rows form a the set of linearly independent vectors in the row-space of $G^*$. Let $G_1$ have dimensions $n(r-t)\times n$. We show this in Fig. \ref{fig:hstar}. Let $P$ be the matrix whose rows are the rest of the independent vectors in $\{0,1\}^{n}$. Let $H^{*}$ be defined as shown in the figure, and let $H^{*}$ be the parity check matrix of an LDPC code $C^{*}$. $H^*$ has dimensions $n(1-t)\times n$. Hence, $C^*$ has rate $t$. It can be seen that, if $G$ corresponds to a Tanner graph with degree distribution pair ($\lambda_G,\rho_G$), and $G_1$ corresponds to a Tanner graph with degree distribution pair ($\lambda_{G_1},\rho_{G_1}$), then $H^{*}$ corresponds to a Tanner graph with degree distribution pair ($\lambda_{H^{*}},\rho_{H^{*}}$), where
\begin{equation}
\frac{\lambda_{H^*}(x)}{\int_{0}^{1}{\lambda_{H^*}(x)dx}}=\frac{1}{\int_{0}^{1}{\lambda_G(x)dx}\int_{0}^{1}{\lambda_{G_1}(x)dx}}\left(\lambda_G(x)\int{\lambda_{G_1}(x)dx}+\lambda_{G_1}(x)\int{\lambda_G(x)dx}\right),
\label{eqn:oldl}
\end{equation}
and
\begin{equation}
\rho_{H^*}(x)=\frac{\int_{0}^{1}{\lambda_{G}(x)dx}}{\int_{0}^{1}{\lambda_{G}(x)dx}+\int_{0}^{1}{\lambda_{G_1}(x)dx}}\rho_{G_1}(x)+\frac{\int_{0}^{1}{\lambda_{G_1}(x)dx}}{\int_{0}^{1}{\lambda_{G}(x)dx}+\int_{0}^{1}{\lambda_{G_1}(x)dx}}\rho_G(x).
\label{eqn:oldr}
\end{equation}

See Appendix \ref{app:proof3} for a proof of the above relations. We restrict the transmitted word $\mathbf{X}$ to be a linear combination of the rows in only $G$ and $G_1$, i.e. all the vectors in $P$ are multiplied by zero. The secrecy code rate now falls to $r-t$. It is important to note that this new secrecy code will have the same security properties as the original code, since only the matrix $G$ determines the security properties of the secrecy code.

\begin{figure}[h]
\begin{center}
\scalebox{0.7}{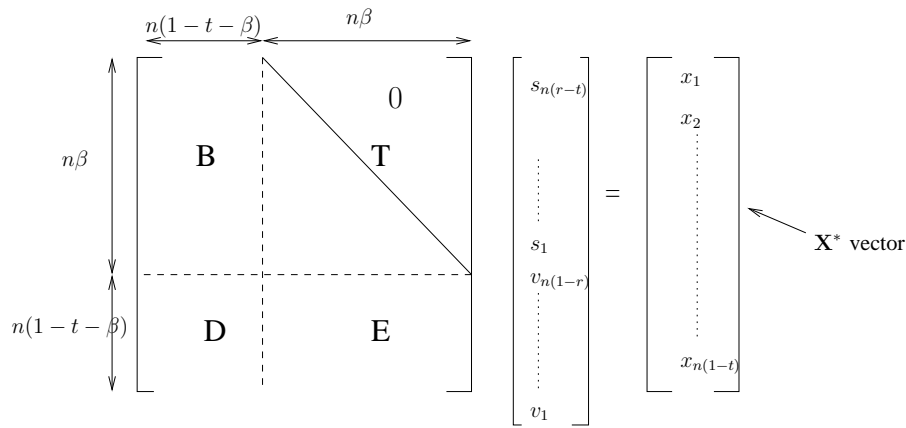}
\end{center}
\caption{The matrix $H$}
\label{fig:hmat1}
\end{figure}

\subsubsection{Forming the matrix $H_1$}
Let the code $C^*$ have erasure threshold $\beta$ under the standard iterative erasure-decoding algorithm. Hence, any submatrix formed using a set of $n\beta$ columns of the $n\times n$ matrix in Fig. \ref{fig:hstar} (i.e. including $G$, $G_1$ and $P$) will have full column rank (asymptotically). By performing some row and column permutations in $G$, $G_1$ and $P$, we can get an approximately upper triangular form in $H^*$. Note that, after row and column permutations, we need to rearrange the $v$-bits, $s$-bits and the $x$-bits. To keep the notation simple we will still call the first $n(1-r)$ bits $v$-bits and the next $n(r-t)$ bits as $s$-bits with the understanding that Bob now will possibly have to find some or all of the bits in not only $\mathbf{S}$, but also in $\mathbf{V}$. After the row and column permutations we continue to call the matrices $G$, $G_1$ and $P$ by the same names.

Now, consider Fig. \ref{fig:hbta}. The matrix $H_1$ is obtained by retaining the $n\beta$ columns in the approximate upper triangular form and by choosing $n(1-t-\beta)$ other columns in such a way that $H_1$ has full column rank in the column-space of $G$ and $G_1$. Thus, $H_1$ will have full column rank in the full $n\times n$ matrix ($G$, $G_1$, and $P$) as well.

\subsubsection{Forming the matrix $H$}
In Fig. \ref{fig:hmat1}, we show the matrix $H$, which is obtained by rotating the matrix $H_1$ in Fig. \ref{fig:hbta} by 90 degrees clockwise. The equation (shown in Fig. \ref{fig:hmat1})
$$H[s_{n(r-t)}\cdots s_1\;v_{n(1-r)}\cdots v_1]^T=[x_1\;x_2\cdots x_{n(1-t)}]^T$$
needs to be solved to find the $s$-bits and the $v$-bits. Note that we have retained only $n(1-t)$ bits in $\mathbf{X}$ on the RHS in Fig. \ref{fig:hmat1}. Since $H$ has full row rank (because $H_1$ in the previous section had full column rank), $n(1-t)$ $x$-bits are enough to solve for the $v$-bits and the $s$-bits. We denote this new vector on the RHS in Fig. \ref{fig:hmat1} as ${\mathbf{X}^*}^T$. We remark that the matrix $H$ is neither the generator matrix of the code $C$ nor the parity check matrix of $C^{\perp}$. 

We will now basically follow the steps described in \cite{effenc} for efficiently solving the equation in Fig. \ref{fig:hmat1}. The matrix $H$ can be divided into matrices $B$, $T$, $D$ and $E$ as in \cite{effenc} (the matrices called $A$ and $C$ in \cite{effenc} are not necessary in our solution) with dimensions $n\beta\times n(1-t-\beta)$, $n\beta \times n\beta$, $n(1-t-\beta) \times n(1-t-\beta)$ and $n(1-t-\beta) \times n\beta$, respectively. $T$ is a lower triangular matrix.

\begin{figure}[h]
\centering
\scalebox{0.7}{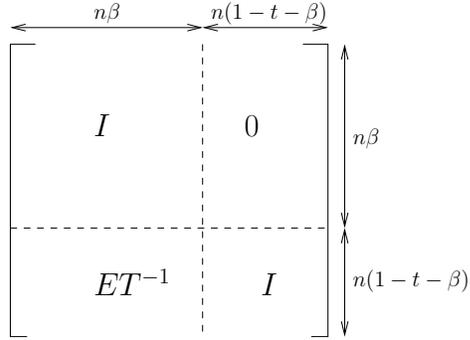}
\caption{The matrix $Q$}
\label{fig:qmat}
\end{figure}
\begin{figure}
\centering
\scalebox{0.7}{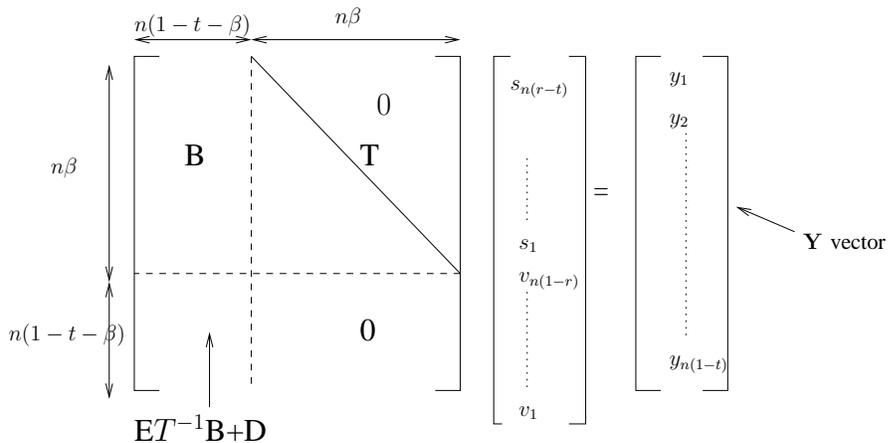}
\caption{The vector $\mathbf{Y}$}
\label{fig:yvect}
\end{figure}

\subsubsection{Multiplying by the matrix Q}
As in \cite{effenc}, we multiply both sides in Fig. \ref{fig:hmat1} by the matrix $Q$ shown in Fig. \ref{fig:qmat}. The result is shown in Fig. \ref{fig:yvect}. The result of the multiplication of the matrix $Q$ with the matrix $H$ can be precomputed before the actual decoding begins. We now study the matrix-vector multiplication $\mathbf{Y}=Q{\mathbf{X}^*}^T$. We need to show that this multiplication is $O(n)$ since we need to do this operation for every received $\mathbf{X}$. 

\begin{figure}[h]
\centering
\scalebox{0.7}{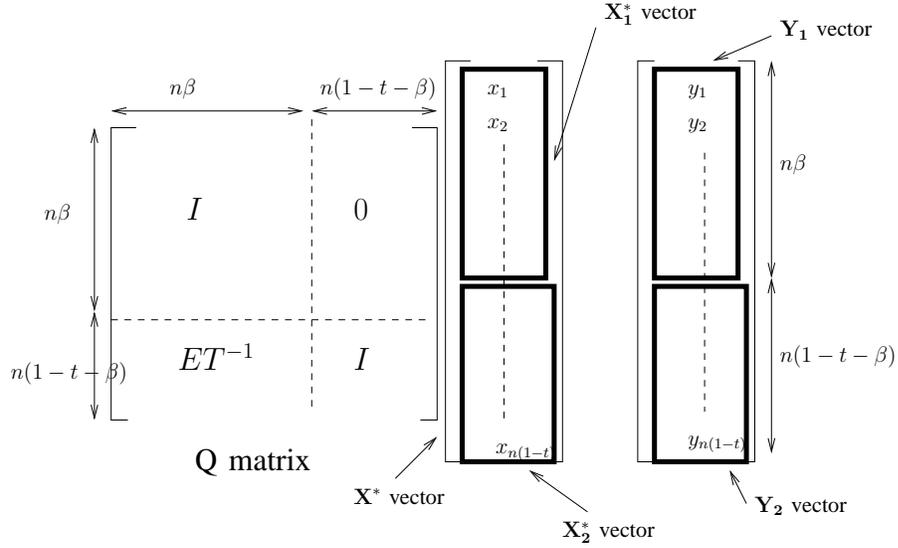}
\caption{The multiplication of Q and $\mathbf{X^*}$}
\label{fig:y12}
\end{figure}

In Fig. \ref{fig:y12}, we have shown the splitting of the vectors $\mathbf{X}^*$ and $\mathbf{Y}$ into vectors $\mathbf{X}^*_1$, $\mathbf{X}^*_2$ and $\mathbf{Y}_1$, $\mathbf{Y}_2$, respectively. The vectors $\mathbf{X}^*_1$ and $\mathbf{Y}_1$ have dimension $1\times n\beta$, while vectors $\mathbf{X}^*_2$ and $\mathbf{Y}_2$ have dimensions $1\times n(1-t-\beta)$. From Fig. \ref{fig:y12}, the vectors $\mathbf{Y}_1$ and $\mathbf{X}^*_1$ are equal. Hence,  $\mathbf{Y}_1$ can be computed in linear time. Now 
$$\mathbf{Y}^T_2=ET^{-1}{\mathbf{X}^*_1}^T+{\mathbf{X}^*_2}^T.$$ 
Clearly, $T^{-1}{\mathbf{X}^*_1}^T$ can be computed using back-substitution in $O(n)$ time, and the multiplication of this result with $E$ and the addition with ${\mathbf{X}^*2}^T$ are also linear time. Hence, the computation of $\mathbf{Y}$ from $Q$ and $\mathbf{X}^*$ is $O(n)$.
\subsubsection{Solving for vectors $\mathbf{S}$ and $\mathbf{V}$}
\label{section:sv}
\begin{figure}
\centering
\scalebox{0.7}{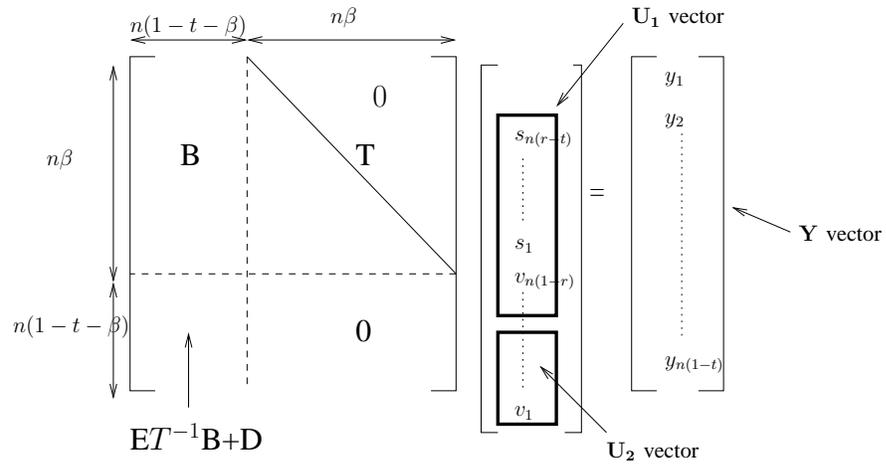}
\caption{Solving for $\mathbf{U_1}$ and $\mathbf{U_2}$}
\label{fig:u12}
\end{figure}

We now turn to Fig. \ref{fig:yvect}. Let the first $n(1-t-\beta)$ elements of the vector $[s_{n(r-t)}\cdots s_1\;v_{n(1-r)}\cdots v_1]$ in the LHS in Fig. \ref{fig:yvect} be denoted by $\mathbf{U}_1$, and the next $n\beta$ elements be denoted by $\mathbf{U}_2$ as shown in Fig. \ref{fig:u12}. 
To compute $\mathbf{S}$ and $\mathbf{V}$, we now need to solve for $\mathbf{U}_1$ and $\mathbf{U}_2$ using
\begin{equation}
B{\mathbf{U}_1}^T+T{\mathbf{U}_2}^T={\mathbf{Y}_1}^T,\text{ and }
\label{eqn:bu1eqn}
\end{equation}
$$(ET^{-1}B+D){\mathbf{U}_1}^T={\mathbf{Y}_2}^T,$$
where $\mathbf{Y}_1$ and $\mathbf{Y}_2$ are as defined before (in Fig. \ref{fig:y12}). Solving the second equation first, we get
$$\mathbf{U}_1=(ET^{-1}B+D)^{-1}\mathbf{Y}_2.$$
Though the inverse can be precomputed, the multiplication is not $O(n)$ (in general), since $(ET^{-1}B+D)^{-1}$ is not sparse anymore. The complexity of this multiplication is $O((1-t-\beta)^2n^2)$. However, if $\beta=1-t$, the vector $\mathbf{U}_1$ is empty and does not have to be computed. From (\ref{eqn:bu1eqn}), 
$$T\mathbf{U}_2=\mathbf{Y}_1+B\mathbf{U}_1.$$ 
Since $B$ is a sparse matrix, $B\mathbf{U}_1$ can be computed in linear time, and then $\mathbf{U}_2$ can be solved in $O(n)$ time by back-substitution, since $T$ is a sparse lower triangular matrix. 

If $\beta=1-t$ the complexity of the entire decoding operation reduces to $O(n)$. Hence, if the code $C^*$ is a capacity-achieving erasure-correcting code, then linear time decoding is possible. As we will discuss in the next section, this is a sufficient condition but not necessary. We now conclude this section with an example of a secrecy code decoded using the method described in this section.
\begin{example}
Let $C$ be a $(3,6)$-regular LDPC code with block-length $n$. Let $G$ be the sparse parity check matrix of $C$ (i.e. the generator matrix of $C^{\perp}$). The rate of $C$ is $r=1/2$. The matrix $G_1$ is chosen to be the parity check matrix of a $(2,6)$-regular LDPC code. Then the code $C^*$ is an LDPC code with all variable nodes having degree $5$ and all check nodes having degree $6$. The rate of $C^*$ is $t=1/6$. The LDPC code $C$ has an erasure threshold $\alpha\approx0.42$. The code $C^*$ has an erasure threshold $\beta\approx0.55$. Thus, the secrecy rate is $r-t=1/3$, and $1-t-\beta\approx0.283$. The decoding complexity is $O(n^2)$ because of a multiplication in the decoder by a $0.283n\times 0.283n$ non-sparse matrix; however, this multiplication is still less complex than a direct decoder that would require a $n\times n$ non-sparse matrix multiplication. The transmitted message is secure across the wiretap channel having erasure probability at least $(1-\alpha)=0.58$. 
\end{example}
\subsection{Linear time decodable secrecy codes}
We now discuss linear-time decodable secrecy codes. As we saw in the previous sections, a sufficient condition for the secrecy code to be linear-time decodable is that the code $C^*$ in Fig. \ref{fig:hstar} (with parity check matrix $H^*$) should be a capacity achieving code on a binary erasure channel so that the row gap in $H^*$ (i.e. $(1-t-\beta)$) is zero. The authors in \cite{effenc} have shown that the row gap can also be calculated using the erasure threshold of the transpose of the parity check matrix ($H^*$). The transpose of $H^*$ does not correspond to a non-zero rate code. Nevertheless, the greedy algorithm that is used to get approximate triangulation in $H^*$ can also be thought to be the standard iterative erasure decoding algorithm operating on the transpose of $H^*$. Let $H^*$ have degree distribution $(\lambda_{H^*},\rho_{H^*})$, where $\lambda_{H^*}$ and $\rho_{H^*}$ correspond to the variable and the check nodes, respectively. Then, the transpose of $H^*$ will have a degree distribution $(\rho_{H^*},\lambda_{H^*})$. In \cite{effenc}, it is shown that the row gap obtained is then $(1-t-\delta)$, where $\delta$ is the erasure threshold of the degree distribution pair $(\rho_{H^*},\lambda_{H^*})$. It turns out that, many of the known degree distributions of good LDPC error correcting codes actually allow linear time encoding (i.e. $(1-t-\delta)$ is zero).

In the following example, we will use this idea to construct a secrecy code that allows linear-time decoding (and encoding since $G$, $G_1$ in Fig. \ref{fig:hstar} are sparse anyway). Since some matrices in our example have a few degree-zero variable nodes, we will use node-based degree distributions as opposed to the typical edge-based degree distribution. Given an edge-based degree distribution $(\lambda(x),\rho(x))$, let $v(x)$ denote the node-based degree distribution of the variable nodes. The coefficient of $x^i$ in $v(x)$, denoted $v_i$, is the fraction of variable nodes with degree $i$. Clearly,
\begin{equation}
v(x)=\frac{\int{\lambda(x)dx}}{\int_{0}^{1}{\lambda(x)dx}}.
\end{equation}
\begin{example}
We refer to Fig. \ref{fig:hstar} for this example. Let the code $C$ (with parity check matrix $G$) have degree distributions $(\lambda_G,\rho_G)$, where $\lambda_G(x)=0.6087x+0.3913x^2$, and $\rho_G(x)=x^6$. Thus, the variable degree distribution, $v_G(x)=0.7x^2+0.3x^3$. Let the matrix $G_1$ correspond to the parity check matrix of a code with degree distributions $(v_{G_1},\rho_{G_1})$, where $v_{G_1}(x)=0.7+0.3x$, and $\rho_{G_1}(x)=x^6$. Hence, the degree distributions of $H^*$ is $(\lambda_{H^*},\rho_{H^*})$, where $\lambda_{H^*}(x)=0.3769x+0.4846x^2+0.1385x^3$, and $\rho_{H^*}(x)=x^6$ (using (\ref{eqn:oldl}) and (\ref{eqn:oldr})). The rate of the secrecy code then is $r(\lambda_G,\rho_G)-r(v_{H^*},\rho_{H^*})=0.0429$, where $r$ denotes the rate of the corresponding LDPC code. Hence, our secrecy rate has dropped to $0.0429$ from $1-(1-r(\lambda_G,\rho_G))=0.6714$. The erasure threshold of $C$ turns out to be $0.2625$. Hence this code is secure on a wiretap channel with erasure probability at least $(1-0.2625)=0.7375$ (i.e. secrecy capacity is $0.7375$). This secrecy code is linear time decodable.
\end{example}
\section{Erasure main channel and Erasure wiretapper's channel}
\label{sec:codmet2}
In this section, we consider wire tap systems where both the wire tap channel and the main channel are binary erasure channels (BEC). Though our results apply with a small modification to systems with DMCs other than the BEC as the main channel, we restrict ourselves to the BEC case for ease of explanation. 

\begin{figure}[htb]
\begin{center}
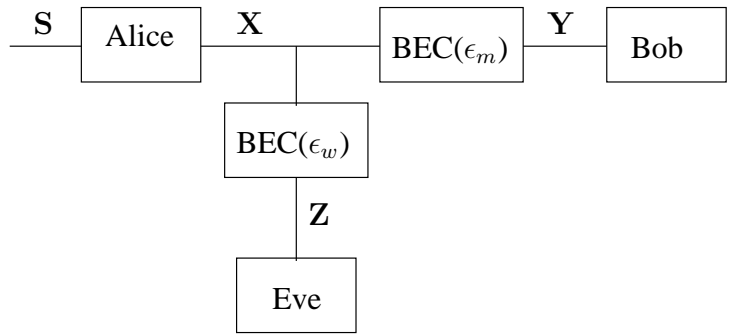
\end{center}
\caption{The BEC wire tap system}
\label{fig:gen_bec}
\end{figure}

With a BEC as the main channel, the wire tap system is as shown in Figure \ref{fig:gen_bec}. 
The wiretapper's channel is a BEC with erasure probability $\epsilon_w$, and the main channel is another BEC with erasure probability $\epsilon_m$. According to (\ref{eqn:Cssym}), the secrecy capacity of this system is $C_s=\epsilon_w-\epsilon_m$, which is positive whenever $\epsilon_w>\epsilon_m$.
\subsection{Using duals of codes on graphs}
As in the noiseless main channel case, we consider using the dual of an LDPC code as the code $C$ for encoding. Using Theorem \ref{thm:ldpc}, security across the wiretapper's channel can be related to the threshold $\alpha$ of the LDPC code $C^{\perp}$ over erasure channels. Specifically, if $1-\epsilon_w<\alpha$, security is guaranteed with high probability.

We now turn to the probability of error on the main channel. Suppose we could design the matrix $G^*$ such that the overall code $\overline{C}$ still belongs to an LDPC ensemble with threshold $\beta$ over erasure channels. Bob can decode $\mathbf{x}$ (and hence the message $\mathbf{s}$) with asymptotically zero probability of error whenever $\epsilon_m<\beta$. 

In summary, the requirement on the LDPC code $\overline{C}$ is that it should contain $C$, the dual of another LDPC code $C^{\perp}$. Since the dual of an LDPC code is likely to have a significantly high number of low-weight codewords, the requirement appears to be contrary to intuition. A very similar code design problem arises in the construction of quantum error-correcting codes using sparse graphs \cite{mackayqecc}. After studying several constructions, the authors of \cite{mackayqecc} conclude that such codes are difficult to construct and are unlikely to have high thresholds.
\subsection{Using capacity-achieving codes}
\label{section:encodingsection}
We now consider a coding method that will eventually depend on capacity-achieving codes for complete security. We first pick an LDPC code $C_1$ of length $n$ from an ensemble of codes having asymptotic erasure threshold $\epsilon_w$. That means, as $n\rightarrow\infty$, $C_1$ recovers all the erasures on an erasure channel with erasure probability up to at least $\epsilon_w$ using the standard iterative erasure decoding algorithm. Let the rate of $C_1$ be $r_1$, and let $H_1$ be the parity check matrix of the code $C_1$. Next we select $n(1-r_2)$ independent vectors from the dual space of $C_1$, where $r_1<r_2$. The selection is such that the $n(1-r_2)\times n$ matrix $H_2$ formed by these vectors as rows is from an LDPC ensemble with erasure threshold $\epsilon_m$. Let $\overline{H}_2$ be the matrix whose rows are the rest of the independent vectors in the dual space of $C_1$. As we will see shortly, we must have $\epsilon_w>(1-r_2)$ in order to guarantee some equivocation for Eve. Let $C_2$ be the code with parity check matrix $H_2$. From capacity considerations, we have
$$1-r_2\geq\epsilon_m,\text{ and}$$
$$1-r_1\geq\epsilon_w.$$
In our examples, we will construct $H_2$ by picking $n(1-r_2)$ rows of $H_1$, and the rest of the rows will be in $\overline{H}_2$. Let $H_2$ correspond to a Tanner graph with degree distribution pair ($\lambda_2,\rho_2$), and let $\overline{H}_2$ correspond to a Tanner graph with degree distribution pair ($\overline{\lambda}_2,\overline{\rho}_2$). Using (\ref{eqn:oldl}) and (\ref{eqn:oldr}), we see that $H_1$ corresponds to a Tanner graph with degree distribution pair ($\lambda_1,\rho_1$), where
\begin{equation}
\frac{\lambda_1(x)}{\int_{0}^{1}{\lambda_1(x)dx}}=\frac{1}{\int_{0}^{1}{\lambda_2(x)dx}\int_{0}^{1}{\overline{\lambda}_2(x)dx}}\left(\lambda_2(x)\int{\overline{\lambda}_2(x)dx}+\overline{\lambda}_2(x)\int{\lambda_2(x)dx}\right),
\label{eqn:newl}
\end{equation}
and
\begin{equation}
\rho_1(x)=\frac{\int_{0}^{1}{\overline{\lambda}_{2}(x)dx}}{\int_{0}^{1}{\lambda_{2}(x)dx}+\int_{0}^{1}{\overline{\lambda}_{2}(x)dx}}\rho_2(x)+\frac{\int_{0}^{1}{\lambda_{2}(x)dx}}{\int_{0}^{1}{\lambda_{2}(x)dx}+\int_{0}^{1}{\overline{\lambda}_{2}(x)dx}}\overline{\rho}_2(x).
\label{eqn:newr}
\end{equation}
We have to choose $(\lambda_1,\rho_1$) and ($\lambda_2,\rho_2$) in such a way that for all $i$, $\overline{\lambda}_{2i}$ and $\overline{\rho}_{2i}$ are non-negative.
\begin{figure}[htb]
\centering
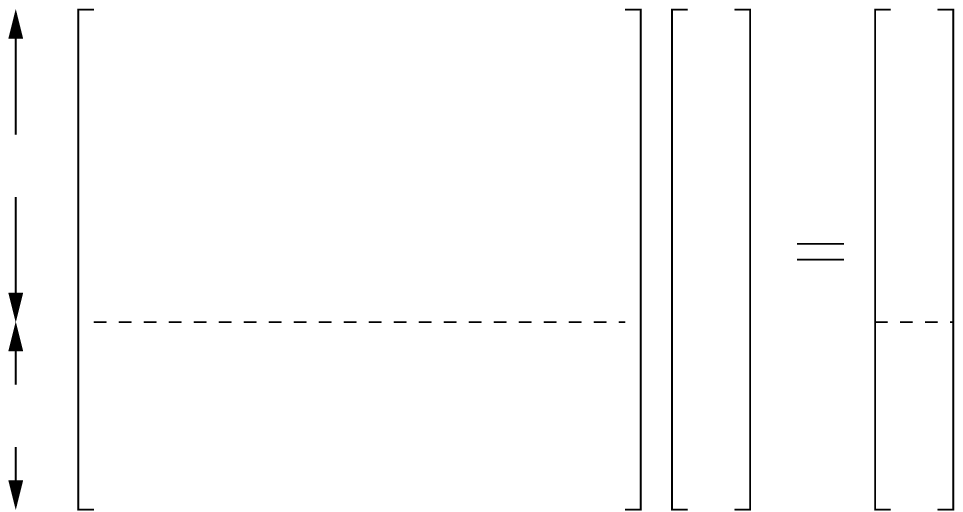
\caption{The encoding procedure}
\label{fig:Hmat}
\end{figure}
\subsubsection{Encoding procedure}
We now discuss the encoding procedure, which is a little different from the encoding procedure for a noiseless main channel. Here, Alice starts with a $n(r_2-r_1)$-bit message vector $\mathbf{S}$, and forms a $n(1-r_1)$-bit vector $[0\cdots0\;\mathbf{S}]$ by prefixing $n(1-r_2)$ $0$'s to $\mathbf{S}$. She now chooses, for transmission, a vector $\mathbf{X}$ at random from the solution set of the equation
\begin{equation}
\begin{bmatrix}H_2\\ \overline{H}_2\end{bmatrix}\mathbf{X}^T=[0\cdots0\;S]^T.
\label{eqn:ext1}
\end{equation}
\begin{figure}[t]
\begin{center}
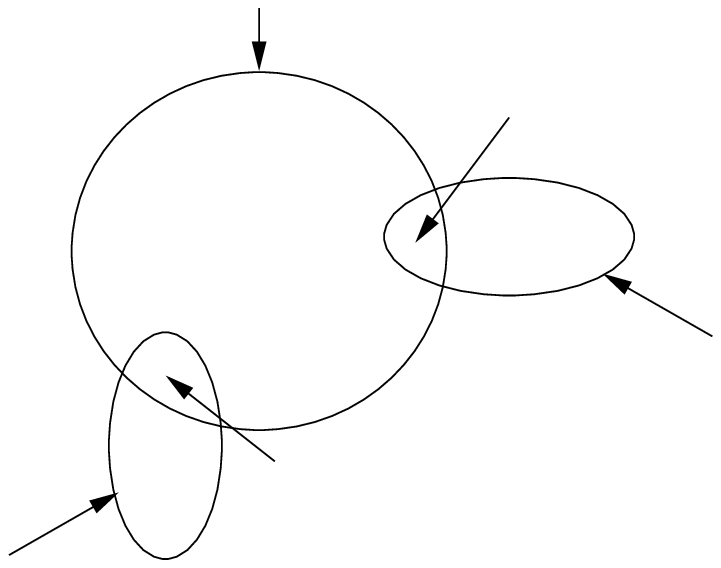
\end{center}
\caption{The encoding space}
\label{fig:enc_space}
\end{figure}
We illustrate this encoding procedure in Fig. \ref{fig:Hmat}. The number of solutions to the equation $H_2\mathbf{X}^T=\mathbf{0}$ is $2^{n-n(1-r_2)}=2^{nr_2}$. For a particular choice of $\mathbf{S}=\mathbf{S}_1$, the number of solutions to (\ref{eqn:ext1}) is $2^{n-n(1-r_1)}=2^{nr_1}$. In addition, the solution sets of (\ref{eqn:ext1}) for different values of $\mathbf{S}$ are disjoint as shown in Fig. \ref{fig:enc_space}. Therefore, the solution space of the equation $H_2\mathbf{X}^T=\mathbf{0}$ splits into $\frac{2^{nr_2}}{2^{nr_1}}=2^{n(r_2-r_1)}$ disjoint subsets, each corresponding to a different value of $\mathbf{S}$. Hence, the rate of the secrecy code is $(r_2-r_1)$. An interesting observation in the encoding process is that we are not using the entire space $\{0,1\}^n$ as in the previous sections where the main channel was noiseless.
\subsubsection{Equivocation across the wire tap channel}
\label{section:securitysection}
In this section, we calculate the equivocation for Eve. Since Eve's channel is a BEC with erasure probability $\epsilon_w$, with probability tending to 1, Eve will have $n\epsilon_w$ erasures as $n\rightarrow \infty$. If we have $\epsilon_w>(1-r_2)$, using $H_2\mathbf{X}^T=\mathbf{0}$, Eve must have at least $2^{n(\epsilon_w-(1-r_2))}$ solutions for $\mathbf{X}$, all of which are equally likely. All these solutions will differ from each other in the erased positions. Since $\epsilon_w$ is the erasure threshold of the code having $H_1$ as the parity-check matrix, any submatrix formed using $n\epsilon_w$ columns of $H_1$ will have full column rank \cite{effenc}. Thus every solution of $H_2\mathbf{X}^T=\mathbf{0}$ will result in a different value of $\mathbf{S}$, all of which are equally likely. The equivocation for Eve is then $\Delta=n(\epsilon_w-(1-r_2))$. If $H_1$ is the parity-check matrix of a capacity-achieving code on an erasure channel with erasure probability $\epsilon_w$, $\Delta=n(r_2-r_1)$, and the message will be completely secure from Eve. Clearly, if the erasure probability of Eve's channel goes up, Eve will still have at least this much equivocation.
\subsubsection{Probability of error on the main channel}
When Bob receives a vector $\mathbf{Y}$, he first decodes it by using the standard iterative erasure decoding technique for LDPC codes on the Tanner graph of the code $C_2$. Let the erasure probability of the main channel be at most $\epsilon_m$. Then, as $n\rightarrow\infty$, with probability tending to $1$ he will be able to recover the transmitted word $\mathbf{X}$. Bob then can find out the product $\overline{H}_2\mathbf{X}^T$, which is his estimate of the message $\mathbf{S}$.

We now illustrate the codes involved in this coding method with an example.
\begin{example}
Let $C_2$ be a $(3,6)$-regular LDPC code with block-length $n$. Hence, $\lambda_2(x)=x^2$ and $\rho_2(x)=x^5$. $C_2$ has rate $r_2=1/2$. The code $C_1$ is chosen to be another LDPC code with all variable nodes having degree $5$ and all check nodes having degree $6$. Hence, $\lambda_1(x)=x^4$ and $\rho_1(x)=x^5$. $C_1$ has rate $r_1=1/6$. It can be seen from (\ref{eqn:newl}) and (\ref{eqn:newr}) that, $\overline{\lambda}_2(x)=x$ and $\overline{\rho}_2(x)=x^5$. The LDPC code $C_2$ has an erasure threshold $\alpha^*\approx0.42$. The code $C_1$ has an erasure threshold $\beta^*\approx0.55$. Thus, the secrecy rate is $r_2-r_1=1/3$, and an equivocation of $n(\beta^*-(1-r_2))=0.05n$ is guaranteed across the wiretap channel having erasure probability greater than $\beta^*=0.55$. Bob can decode the message with asymptotically zero probability of error on the main channel having erasure probability at most $\alpha^*=0.42$.
\end{example}
The example above illustrates the main drawback of this coding method. Unless the code $C_1$ is capacity-achieving ($\beta^*=1-r_1$), the coding method is not secure even for large $n$. Note that the equivocation could be reduced further by a better choice of $C_1$, but the equivocation will go to zero only for capacity-achieving codes. 
\subsection{Remarks}
We have shown that codes on graphs provide secrecy in erasure wire tap channels with maximum possible secure information rate. The codes are efficiently implementable in practice. However, if the main channel is not noiseless, secrecy by forward coding alone appears to require capacity-achieving codes that are difficult to construct. Alternative models of wire tap channels with parallel error-free public channels are presumably better for constructing implementable secrecy codes when the main channel is noisy.
\section{Noiseless main channel and BSC wiretapper's channel}
\label{sec:BSC}
In this section, we consider a special case of a wire tap channel, where the eavesdropper sees a binary symmetric channel (BSC) with error probability $p$, denoted BSC($p$). The main channel is error free. Using (\ref{eqn:Cssym}), we see that
\begin{equation}
\text{C}_s=1-\text{Capacity}(\text{BSC}(p))=1-(1-h(p))=h(p),
\label{eqn:CsBSC}
\end{equation}
where $h(x)=-x\log_2x-(1-x)\log_2(1-x),\;0\leq x\leq1$.

\begin{figure}[ht]
\begin{center}
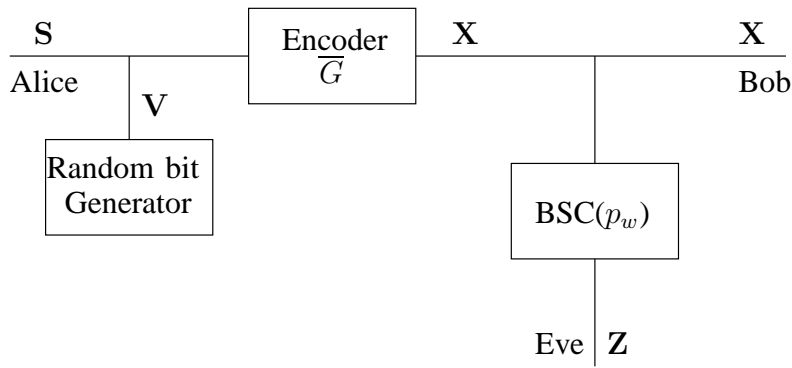
\end{center}
\caption{Coding for a BSC wiretapper's channel.}
\label{fig:BSCWTC}
\end{figure}

The wire tap channel and the encoding is shown in Fig. \ref{fig:BSCWTC}.
The method of coding is illustrated with the same notation as Section \ref{sec:BEC}. 
\subsection{Security across a BSC wiretapper's channel}
We let $C$ be an $(n,n-k)$ code and $\overline{C}$ be the entire space $\{0,1\}^n$. For an arbitrary $k$-bit message $\mathbf{S}=\mathbf{s}$, the transmitted word $\mathbf{X}\in \mathbf{s}G^*+C$. Since the cosets of $C$ cover the entire space of $n$-tuples, Eve's received vector $Z$ belongs to some coset of $C$, say $\mathbf{u}G^*+C$. If $\mathbf{e}$ denotes the error vector introduced by the BSC($p$) in the wiretap, we have for $1\leq i\leq 2^k$,
\begin{equation}
\text{Prob}\{Z\in \mathbf{u}G^*+C|\mathbf{S}=\mathbf{s}\}=\text{Prob}\{\mathbf{e}\in(\mathbf{u}+\mathbf{s})G^*+C\}=\text{Prob}\{e\in \mathbf{w}+C\}\text{ for some $n$-tuple }\mathbf{w}.
\label{eqn:Ze}
\end{equation}
We can now state the criterion for selecting the code $C$ to guarantee security of the message $\mathbf{S}$: we choose $C$ such that for any $n$-tuple $\mathbf{w}$, we have
\begin{equation}
\text{Prob}\{e\in \mathbf{w}+C\}\rightarrow2^{-k}, \text{ as }n\rightarrow\infty.
\label{eqn:w}
\end{equation}
Using the above condition in (\ref{eqn:Ze}), we see that Eve is equally likely to find $\mathbf{Z}$ in any coset of $C$ given any message $\mathbf{S}=\mathbf{s}$. Assuming all $\mathbf{S}=\mathbf{s}$ are equally likely {\it{a priori}}, Prob$\{\mathbf{Z}\in \mathbf{u}G^*+C\}$ is independent of $\mathbf{u}$; hence, Prob$\{\mathbf{S}=\mathbf{s}|\mathbf{Z}\in \mathbf{u}G^*+C\}\rightarrow2^{-k}$, and security is guaranteed.

The LHS of (\ref{eqn:w}) is the probability of the coset $\mathbf{w}+C$. This probability was first studied by Sullivan \cite{Sullivan} and further extended by Ancheta \cite{Ancheta1, Ancheta2}. The following results can be extracted from their studies: (1) The requirement of (\ref{eqn:w}) that the probabilities of a code ($w=0$) and a coset ($w\ne0$) should be approximately equal can be achieved for large block-length. (2) The properties of the dual of a code plays an important role in the probability of a coset. We expand on these two results in the next sections to design codes for the BSC wire tap channel.
\subsection{Choosing the code $C$: Security criterion}
Using the MacWilliams identities \cite[Page 127]{MacWilliams:book} for the $(n,n-k)$ linear code $C$, we get
\begin{equation}
\sum_{e\in C}x^{n-\text{wt}(e)}y^{\text{wt}(e)}=\frac{1}{2^k}\sum_{i=0}^{n}A'_i(x+y)^{n-i}(x-y)^i,
\label{eqn:mwil}
\end{equation}
where $A'_i$ is the number of codewords of weight $i$ in the dual code $C^{\perp}_2$. Using $x=1-p$, $y=p$, and $A'_0=1$ in (\ref{eqn:mwil}), we get
$$\sum_{e\in C}p^{\text{wt}(e)}(1-p)^{n-\text{wt}(e)}=2^{-k}+2^{-k}\sum_{i=1}^{n}A'_i(1-2p)^i.$$
Using the MacWilliams identities \cite[Page 137]{MacWilliams:book} for the coset $\mathbf{w}+C$, we get
\begin{equation}
\sum_{e\in \mathbf{w}+C}x^{n-\text{wt}(e)}y^{\text{wt}(e)}=\frac{1}{2^k}\sum_{i=0}^{n}A'_i(\mathbf{w})(x+y)^{n-i}(x-y)^i,
\label{eqn:mwinl}
\end{equation}
where
\begin{equation}
A'_i(\mathbf{w})=\alpha_i(\mathbf{w})-\beta_i(\mathbf{w})
\label{eqn:Apw}
\end{equation}
with $\alpha_i(\mathbf{w})$ equal to the number of codewords of weight $i$ in the dual code $C^{\perp}_2$ orthogonal to $\mathbf{w}$, and $\beta_i(\mathbf{w})$ equal to the number of codewords of weight $i$ in the dual code $C^{\perp}_2$ not orthogonal to $\mathbf{w}$. Using $x=1-p$, $y=p$, and $A'_0(\mathbf{w})=1$ in (\ref{eqn:mwinl}), we get
\begin{equation}
\sum_{e\in \mathbf{w}+C}p^{\text{wt}(e)}(1-p)^{n-\text{wt}(e)}=2^{-k}+2^{-k}\sum_{i=1}^{n}A'_i(\mathbf{w})(1-2p)^i.
\label{eqn:Awcond}
\end{equation}
 From (\ref{eqn:Apw}), we see that $|A'_i(\mathbf{w})|\leq A'_i$. We now state the main security criterion as a theorem.
\begin{thm}
If
\begin{equation}
\sum_{i=1}^{n}A'_i(1-2p)^i\rightarrow0,\text{ as }n\rightarrow\infty,
\label{eqn:BSCcrit}
\end{equation}
then $\text{Prob}\{e\in \mathbf{w}+C\}\rightarrow2^{-k}$ for all $n$-tuples $\mathbf{w}$.
\label{thm:BSCcrit}
\end{thm}
\begin{proof}
Since $|A'_i(\mathbf{w})|\leq A'_i$, we get
$$|\sum_{i=1}^{n}A'_i(\mathbf{w})(1-2p)^i|\leq\sum_{i=1}^{n}A'_i(1-2p)^i\rightarrow0.$$
Hence,
$$|\sum_{i=1}^{n}A'_i(\mathbf{w})(1-2p)^i|\rightarrow0.$$
That implies that the second term in the RHS of (\ref{eqn:Awcond}) can be neglected with respect to the first term $2^{-k}$, and the proof is complete.
\end{proof}
The criterion for the selection of $C$ is that the dual $C^{\perp}$ should have a weight distribution that satisfies (\ref{eqn:BSCcrit}).
\subsection{Some code constructions}
We provide some examples of codes that satisfy the requirement of (\ref{eqn:BSCcrit}).
\begin{example} (Single parity check codes) The dual of a $(n,n-1,2)$ single parity check code is the $(n,1,n)$ repetition code with weight distribution $A'_0=1$ and $A'_n=1$. Hence,$$\sum_{i=1}^{n}A'_i(1-2p)^i=(1-2p)^n\rightarrow0$$
as $n\rightarrow\infty$. However, the secrecy rate $1/n\rightarrow0$ for large $n$. This is an example that was first used by Wyner in \cite{Wyner1} to motivate coding over a wire tap channel.
\end{example}
\begin{example}(Hamming codes) The weight distribution of the dual of the $[n=2^m-1,n-m,3]$ Hamming code $\mathcal{H}_m$ is $A'_0=1$ and $A'_{(n+1)/2}=n$. Hence,
$$\sum_{i=1}^{n}A'_i(1-2p)^i=n(1-2p)^{(n+1)/2}\rightarrow0$$
as $n\rightarrow\infty$. As in the previous example, the secrecy rate tends to zero for large $n$.
\end{example}

The following theorem generalizes the above construction method.
\begin{thm}
Let $\{C_{(n)}\}$ be a sequence of $(n,n-k_n)$ codes such that Prob\{Detection Error\}$\leq2^{-k_n}$ over a BSC($p$), $0\leq p\leq1/2$ and $\lim_{n\rightarrow\infty}\{k_n/n\}<\log_2(1/(1-p))$. Let $A'_i$ be the number of codewords of weight $i$ in the dual code $C^{\perp}_{(n)}$. Then
$$\sum_{i=1}^{n}A'_i(1-2p)^i\rightarrow0,\;\text{as }n\rightarrow\infty.$$
\label{thm:C2}
\end{thm}
\begin{proof}
We are given that for the code $C_{(n)}$
$$\text{Prob\{Detection Error\}}=\sum_{e\in C_{(n)};e\ne0}p^{\text{wt}(e)}(1-p)^{n-\text{wt}(e)}\leq2^{-k_n}.$$
Adding $(1-p)^n$ to both sides and using the MacWilliams identities, we get 
$$\sum_{e\in C_{(n)}}p^{\text{wt}(e)}(1-p)^{n-\text{wt}(e)}=2^{-k_n}+2^{-k_n}\sum_{i=1}^{n}A'_i(1-2p)^i\leq (1-p)^n+2^{-k_n},$$
or
$$\sum_{i=1}^{n}A'_i(1-2p)^i\leq 2^{k_n}(1-p)^n=2^{n(k_n/n-\log_2(1/(1-p)))}.$$
Since $\lim_{n\rightarrow\infty}\{k_n/n\}<\log_2(1/(1-p))$ and the LHS above is nonnegative,
$$\sum_{i=1}^{n}A'_i(1-2p)^i\rightarrow0,\;\text{as }n\rightarrow\infty.$$
\end{proof}
The existence of $(n,n-k_n)$ linear codes with probability of detection error less than $2^{-k_n}$ is well known \cite[Section 3.6]{Costello:book}. Suppose we find a class of such error detecting codes such that 
$$R=\lim_{n\rightarrow\infty}\frac{k_n}{n}.$$
Then, for large $n$, the code $C_{(n)}$, when used as the code $C$ over a wire tap channel with a BSC$(p)$ as the wiretapper's channel, provides security whenever $R<-\log_2(1-p)$, or $p>1-2^{-R}$. The maximum possible secrecy rate that can be achieved by this construction is therefore $-\log_2(1-p)$.

Codes such as Hamming codes and double error-correcting BCH codes are examples of such error-detecting codes. However, most known classes of such codes have $R=0$.
\section{Conclusion and Discussion}
\label{sec:conclusion}
In this paper, we have studied the construction of codes that provide security and reliability over a wire tap channel. Our general construction uses codes that approach capacity over the wire tapper's channel. We have shown that this approach achieves secrecy capacity when the wire tap channel is made of symmetric DMCs. Other cases require a closer study.

A drawback of using capacity-achieving codes is that they are difficult to find and construct except in some special cases. One such special case is when the wire tap channel is a binary erasure channel. Hence, codes such as optimized Tornado codes can be used across erasure wiretapper's channels as described above. However, we have shown that capacity-achieving codes are not necessary in this case. If a code exhibits a threshold behavior across a BEC (codes such as regular LDPC codes), its dual can be used effectively over a wire tap channel with a BEC as the wiretapper's channel. This result enables the use of codes that can be more easily constructed. Extending the connections between codes that exhibit a threshold phenomenon and secrecy over a general DMC is an important area of future work.

When the wiretapper's channel is a BEC and the main channel is noiseless, we have presented codes that approach secrecy capacity. To our knowledge these are the first and only such codes.

For the case where both the main channel and the wiretapper's channel are BECs, we have studied two approaches for code design. The optimality and secrecy capacity of the constructions need to be studied and explored.

For the case where the wiretapper's channel is a BSC$(p)$ and the main channel is noiseless, we have shown that codes with good error-detecting properties provide security. The capacity of this construction is $-\log_2(1-p)$, which is less than the secrecy capacity $h(p)$. Capacity-approaching codes will probably be graph-based. Use of graph-based codes for the BSC wiretapper's channel is a subject for future study.
\appendices
\section{Existence of reliable encoders}
\label{app:proof2}
In this section, we determine a random coding bound on the probability of error Prob$\{\mathbf{U}\ne\mathbf{\hat{U}}\}$ in a manner following Gallager \cite[Section 5.6]{Gallager:it}. Let $\mathbf{x}$ be a vector of $n$ input symbols, $\mathbf{y}$ a vector of $n$ main channel output symbols, and $\mathbf{z}$ a vector of $n$ wire tap channel output symbols. We let the output alphabet of the main channel be $\{1,2,\cdots,J_m\}$ denoting a symbol by $j_m$. The eavesdropper on the wire tap channel is assumed to have unlimited power to process the received word $\mathbf{z}$. We let the output alphabet of the wire tap channel be $\{1,2,\cdots,J_w\}$ denoting a symbol by $j_w$. Let $T_n(\mathbf{y}|\mathbf{x})$ and $S_n(\mathbf{z}|\mathbf{x})$ be the transition probabilities for the main channel and wire tap channel, respectively. Let $TS_n(\mathbf{y},\mathbf{z}|\mathbf{x})$ be the joint distribution.

We now define a random code ensemble for the coding method of Section \ref{sec:codmet}. Let $Q_n(\mathbf{x})$ be an arbitrary probability assignment on the set of length $n$ input sequences. A set of $ML$ words is chosen pairwise independently from the set of length $n$ input sequences according to $Q_n(\mathbf{x})$. The words are arranged in an $M\times L$ array indexed by a pair of coordinates $u\in\{1,2,\cdots,M\}$ and $v\in\{1,2,\cdots,L\}$; each word is denoted $\mathbf{x}_m$, where $m=(u,v)$. Each row is considered to be the code $C_{u}$ i.e. $C_{u}=\{\mathbf{x}_{m'}:m'=(u',v');u'=u\}$.

Let us assume that a message $u$ is to be transmitted by Alice. Let us further assume that the word $\mathbf{x}_m$ with $m=(u,v)$ is chosen for transmission from $C_u$. Let $\mathbf{y}$ and $\mathbf{z}$ be the received vectors for Bob and Eve, respectively. We will upper bound the probability of an event $E$, which captures both the security and reliability constraints. The event $E$ is the union of the following two events:
\begin{enumerate}
\item Event $A_{m'}$: an $\mathbf{x}_{m'}$  for $m'=(u',v')\ne m=(u,v)$ with $u\ne u'$ is chosen in the code such that $T_n(\mathbf{y}|\mathbf{x}_{m'})\geq T_n(\mathbf{y}|\mathbf{x}_m)$. This event captures the reliability requirement.
\item Event $B_{m'}$: an $\mathbf{x}_{m'}$ for $m'=(u,v')\ne m=(u,v)$ is chosen in the code such that $S_n(\mathbf{z}|\mathbf{x}_{m'})\geq S_n(\mathbf{z}|\mathbf{x}_m)$. This event captures the security requirement.
\end{enumerate}
The probability of $E$ averaged over the ensemble for the $m=(u,v)$-th word is
\begin{equation}
\overline{P}_{E,m}=\sum_{\mathbf{x}_m}\sum_{\mathbf{y}}\sum_{\mathbf{z}}Q_n(\mathbf{x}_m) TS_n(\mathbf{y},\mathbf{z}|\mathbf{x}_m) \text{Pr}\{E|m,\mathbf{x}_m,\mathbf{y},\mathbf{z}\}
\label{eqn:Pem}
\end{equation}
Using a modified union bound,
$$\text{Pr}\{E|m,\mathbf{x}_m,\mathbf{y},\mathbf{z}\}\leq \left[\sum_{u\ne u'}\text{Pr}\{A_{m'}\}\right]^{\rho_1}+\left[\sum_{u=u',v\ne v'}\text{Pr}\{B_{m'}\}\right]^{\rho_2},$$
for $0\leq\rho_1,\rho_2\leq1$. Now,
\begin{eqnarray*}
\text{Pr}\{A_{m'}\}&=&\sum_{\mathbf{x}_{m'}:T_n(\mathbf{y}|\mathbf{x}_{m'})\geq T_n(\mathbf{y}|\mathbf{x}_m)}Q_n(\mathbf{x}_{m'})\\
&\leq&\sum_{\mathbf{x}}Q_n(\mathbf{x})\frac{T_n(\mathbf{y}|\mathbf{x})^{s_1}}{T_n(\mathbf{y}|\mathbf{x}_m)^{s_1}},\;s_1>0.
\end{eqnarray*}
Similarly,
$$\text{Pr}\{B_{m'}\}\leq\sum_{\mathbf{x}}Q_n(\mathbf{x})\frac{S_n(\mathbf{z}|\mathbf{x})^{s_2}}{S_n(\mathbf{z}|\mathbf{x}_m)^{s_2}},\;s_2>0.$$
Using the simplifications in (\ref{eqn:Pem}),
\begin{eqnarray*}
\overline{P}_{E,m}&\leq&\sum_{\mathbf{x}_m}\sum_{\mathbf{y}}\sum_{\mathbf{z}}Q_n(\mathbf{x}_m) TS_n(\mathbf{y},\mathbf{z}|\mathbf{x}_m)\left[(M-1)L\sum_{\mathbf{x}}Q_n(\mathbf{x})\frac{T_n(\mathbf{y}|\mathbf{x})^{s_1}}{T_n(\mathbf{y}|\mathbf{x}_m)^{s_1}}\right]^{\rho_1}\\
&&+\sum_{\mathbf{x}_m}\sum_{\mathbf{y}}\sum_{\mathbf{z}}Q_n(\mathbf{x}_m) TS_n(\mathbf{y},\mathbf{z}|\mathbf{x}_m)\left[(L-1)\sum_{\mathbf{x}}Q_n(\mathbf{x})\frac{S_n(\mathbf{z}|\mathbf{x})^{s_2}}{S_n(\mathbf{z}|\mathbf{x}_m)^{s_2}}\right]^{\rho_2}\\
&=&(M-1)^{\rho_1}L^{\rho_1}\sum_{\mathbf{y}}\left[\sum_{\mathbf{x}_m}Q_n(\mathbf{x}_m)T_n(\mathbf{y}|\mathbf{x})^{1-s_1\rho_1}\right]\left[\sum_{\mathbf{x}}Q_n(\mathbf{x})T_n(\mathbf{y}|\mathbf{x})^{s_1}\right]^{\rho_1}\\
&&+(L-1)^{\rho_2}\sum_{\mathbf{z}}\left[\sum_{\mathbf{x}_m}Q_n(\mathbf{x}_m)S_n(\mathbf{z}|\mathbf{x})^{1-s_2\rho_2}\right]\left[\sum_{\mathbf{x}}Q_n(\mathbf{x})S_n(\mathbf{z}|\mathbf{x})^{s_2}\right]^{\rho_2}.
\end{eqnarray*}
Using $s_i=1/(1+\rho_i)$, we get a version of Theorem 5.6.1 in Gallager \cite{Gallager:it}. Following Gallager \cite[Section 5.6]{Gallager:it} further for the case of discrete memoryless channels, we let
$$Q_n(\mathbf{x})=\prod_{i=1}^{n}Q(x_i),$$
where the input vector $\mathbf{x}=[x_1\;x_2\cdots x_n]$ in terms of its components, and $Q(k),\;k\in\{1,2,\cdots,K\}$ is an arbitrary probability assignment on the input alphabet. Similarly, we let $T_n(\mathbf{y}|\mathbf{x})=\prod_{i=1}^{n}T(y_i|x_i)$ and $S_n(\mathbf{z}|\mathbf{x})=\prod_{i=1}^{n}S(z_i|x_i)$. Converting to exponential relationships, we get
\begin{equation}
\overline{P}_{E,m}\leq\exp\{-n[E_1(\rho_1,Q)-\rho_1R_1]\}+\exp\{-n[E_2(\rho_2,Q)-\rho_2R_2]\},
\label{eqn:PbarEm}
\end{equation}
where $ML=e^{nR_1}$; $L=e^{nR_2}$;
\begin{eqnarray}
\label{eqn:E1}E_1(\rho_1,Q)&=&-\log\left(\sum_{j_m=1}^{J_m}\left[\sum_{k=1}^{K}Q(k)T(j_m|k)^{1/(1+\rho_1)}\right]^{1|\rho_1}\right);\text{ and}\\
\label{eqn:E2}E_2(\rho_2,Q)&=&-\log\left(\sum_{j_w=1}^{J_w}\left[\sum_{k=1}^{K}Q(k)S(j_w|k)^{1/(1+\rho_2)}\right]^{1|\rho_2}\right).
\end{eqnarray}
Note that the secrecy rate of a code from the ensemble is $R_s=R_1-R_2$. Using a distribution Pr$\{m\}$ in (\ref{eqn:PbarEm}), we get
\begin{equation}
\overline{P}_{E}\leq\exp\{-n[E_1(\rho_1,Q)-\rho_1R_1]\}+\exp\{-n[E_2(\rho_2,Q)-\rho_2R_2]\},
\label{eqn:PbarE}
\end{equation}
The random coding exponent for the wire tap channel is defined as follows:
\begin{equation}
E_w(R_2)=\max_{0\leq\rho_2\leq1}\max_Q[E_2(\rho_2,Q)-\rho_2R_2].
\label{eqn:EwR2}
\end{equation}
Let $Q_2$ be the distribution on the input symbols that maximizes the random coding exponent $E_w(R_2)$. To satisfy the security constraint of Section \ref{sec:seccon}, we restrict ourselves to ensemble of codes with input symbol distribution $Q_2(k)$. We can now define another random coding exponent for the main channel as follows:
$$E_m(R_1)=\max_{0\leq\rho_1\leq1}[E_1(\rho_1,Q_2)-\rho_1R_1].$$
Using the random coding exponents in (\ref{eqn:PbarE}), we get the following theorem.
\begin{thm}
For an ensemble of codes using the maximizing distribution $Q_2$,
\begin{eqnarray*}
\overline{P}_{E,m}&\leq&\exp\{[-nE_m(R_1)]\}+\exp\{[-nE_w(R_2)]\};\\
\overline{P}_{E}&\leq&\exp\{[-nE_m(R_1)]\}+\exp\{[-nE_w(R_2)]\}.
\end{eqnarray*}
\label{thm:relcon}
\end{thm}
We know that $E_w(R_2)>0$ for $0\leq R_2<\text{C}_w$, where $\text{C}_w$ is the channel capacity of the wiretapper's channel. Hence, Theorem \ref{thm:relcon} says that there exists a code in a suitable ensemble such that the security constraint can be satisfied (each $C_u$ can approach capacity on the wire tapper's channel) with arbitrary accuracy by increasing the block-length; at the same time, the same code can satisfy the reliability constraint with arbitrary accuracy provided the rate $R_1$ is such that $E_m(R_1)>0$. From the properties of random coding exponents \cite[Section 5.6]{Gallager:it}, we see that $E_m(R_1)>0$ if
$$R_1<I(Q_2;S)=\sum_{k=1}^{K}\sum_{j=1}^{J_w}Q_2(k)S(j_w|k)\log\frac{S(j_w|k)}{\sum_iQ_2(i)S(j_w|i)}.$$
Hence, the maximum secrecy rate achievable by a code from the ensemble is $I(Q_2;S)-\text{C}_w$. We immediately see that for the special case of a wire tap channel considered in (\ref{eqn:Cssym}) secrecy capacity is achievable by some code in the ensemble. In particular, if both the main channel and wire tapper's channel are symmetric, secrecy capacity is achievable.
\section{Degree Distribution: Proofs for (\ref{eqn:oldl}) and (\ref{eqn:oldr})}
\label{app:proof3}
Let $\lambda_G(x)=\sum_{i}\lambda_ix^{i-1}$, $\rho_G(x)=\sum_{i}\rho_ix^{i-1}$, $\lambda_{G_1}(x)=\sum_{i}\lambda_{1i}x^{i-1}$, $\rho_G(x)=\sum_{i}\rho_{1i}x^{i-1}$, $\lambda_{H^*}(x)=\sum_{i}\lambda_{*i}x^{i-1}$, $\rho_{H^*}(x)=\sum_{i}\rho_{*i}x^{i-1}$.
 
Let $E_1$ and $E_2$ be the total number of 1s in the matrices $G$ and $G_1$, respectively. From the definition of degree distribution, the number of 1s in $G$ and $G_1$ from rows of weight $i$ equals $\rho_iE_1$ and $\rho_{1i}E_2$, respectively. Therefore, $\rho_{*i}$ (fraction of 1s in $H^*$ from rows of weight $i$) is given by
\begin{equation}
\rho_{*i}=\frac{\rho_iE_1+\rho_{1i}E_2}{E_1+E_2}=\frac{E_1}{E_1+E_2}\rho_i+\frac{E_2}{E_1+E_2}\rho_{1i}.
\label{eqn:app1}
\end{equation}
The number of weight $i$ columns in $G$ and $G_1$ equals $\dfrac{\lambda_i}{i}E_1$ and $\dfrac{\lambda_{1i}}{i}E_2$, respectively. The total number of columns in $G$ or $G_1$, $n$, can be written as 
$$n=\sum_i\dfrac{\lambda_i}{i}E_1=\sum_i\dfrac{\lambda_{1i}}{i}E_2.$$
Using the above relations in (\ref{eqn:app1}) for $E_1$ and $E_2$ and replacing $\sum_i\dfrac{\lambda_i}{i}$ (respectively $\sum_i\dfrac{\lambda_{1i}}{i}$) with $\int^1_0\lambda_G(x)dx$ (respectively $\int^1_0\lambda_{G_1}(x)dx)$, we get (\ref{eqn:oldr}).

To prove (\ref{eqn:oldl}), we use the node-based degree distribution of LDPC ensembles. The coefficient of $x^i$ in 
\begin{equation}
\frac{\int^x_{u=0}\lambda_{G}(u)du}{\int^1_0\lambda_G(x)dx}
\label{eqn:app2}
\end{equation}
equals the probability that a randomly chosen column of $G$ has weight $i$. Similarly, the coefficient of $x^i$ in 
\begin{equation}
\frac{\int^x_{u=0}\lambda_{G_1}(u)du}{\int^1_0\lambda_{G_1}(x)dx}
\label{eqn:app3}
\end{equation}
equals the probability that a randomly chosen column of $G_1$ has weight $i$. Note that the polynomials in (\ref{eqn:app2}) and (\ref{eqn:app3}) are generating functions of independent random variables denoting the weight of a randomly chosen column in $G$ and $G_1$, respectively. Since the weight of a column of $H^*$ equals the sum of the weight of the column in $G$ and the weight of the column in $G_1$ and the two weights in $G$ and $G_1$ are independent, we have
$$\frac{\int^x_{u=0}\lambda_{H^*}(u)du}{\int^1_0\lambda_{H^*}(x)dx}=\left(\frac{\int^x_{u=0}\lambda_{G}(u)du}{\int^1_0\lambda_G(x)dx}\right)\left(\frac{\int^x_{u=0}\lambda_{G_1}(u)du}{\int^1_0\lambda_{G_1}(x)dx}\right).$$
Differentiating the above equation gives (\ref{eqn:oldl}).
\bibliographystyle{IEEEtran}
\bibliography{qkd,ldpc,labo}
\end{document}

%% file: broadcast.pstex_t
\begin{picture}(0,0)%
\includegraphics{broadcast}%
\end{picture}%
\setlength{\unitlength}{3947sp}%
\begingroup\makeatletter\ifx\SetFigFont\undefined%
\gdef\SetFigFont#1#2#3#4#5{%
  \reset@font\fontsize{#1}{#2pt}%
  \fontfamily{#3}\fontseries{#4}\fontshape{#5}%
  \selectfont}%
\fi\endgroup%
\begin{picture}(4524,2067)(2089,-2623)
\put(2701,-811){\makebox(0,0)[lb]{\smash{\SetFigFont{12}{14.4}{\familydefault}{\mddefault}{\updefault}{\color[rgb]{0,0,0}Alice}%
}}}
\put(3526,-736){\makebox(0,0)[lb]{\smash{\SetFigFont{12}{14.4}{\familydefault}{\mddefault}{\updefault}{\color[rgb]{0,0,0}$\mathbf{X}$}%
}}}
\put(3751,-2461){\makebox(0,0)[lb]{\smash{\SetFigFont{12}{14.4}{\familydefault}{\mddefault}{\updefault}{\color[rgb]{0,0,0}Eve}%
}}}
\put(3751,-1486){\makebox(0,0)[lb]{\smash{\SetFigFont{12}{14.4}{\familydefault}{\mddefault}{\updefault}{\color[rgb]{0,0,0}C2}%
}}}
\put(4801,-811){\makebox(0,0)[lb]{\smash{\SetFigFont{12}{14.4}{\familydefault}{\mddefault}{\updefault}{\color[rgb]{0,0,0}C1}%
}}}
\put(6001,-886){\makebox(0,0)[lb]{\smash{\SetFigFont{12}{14.4}{\familydefault}{\mddefault}{\updefault}{\color[rgb]{0,0,0}Bob}%
}}}
\put(3976,-1936){\makebox(0,0)[lb]{\smash{\SetFigFont{12}{14.4}{\familydefault}{\mddefault}{\updefault}{\color[rgb]{0,0,0}$\mathbf{Z}$}%
}}}
\put(5326,-736){\makebox(0,0)[lb]{\smash{\SetFigFont{12}{14.4}{\familydefault}{\mddefault}{\updefault}{\color[rgb]{0,0,0}$\mathbf{Y}$}%
}}}
\put(2176,-736){\makebox(0,0)[lb]{\smash{\SetFigFont{12}{14.4}{\familydefault}{\mddefault}{\updefault}{\color[rgb]{0,0,0}$\mathbf{M}$}%
}}}
\end{picture}

%% file: codmet.pstex_t
\begin{picture}(0,0)%
\includegraphics{codmet}%
\end{picture}%
\setlength{\unitlength}{3947sp}%
\begingroup\makeatletter\ifx\SetFigFont\undefined%
\gdef\SetFigFont#1#2#3#4#5{%
  \reset@font\fontsize{#1}{#2pt}%
  \fontfamily{#3}\fontseries{#4}\fontshape{#5}%
  \selectfont}%
\fi\endgroup%
\begin{picture}(7374,2724)(814,-3148)
\put(1726,-961){\makebox(0,0)[lb]{\smash{\SetFigFont{12}{14.4}{\familydefault}{\mddefault}{\updefault}{\color[rgb]{0,0,0}$\mathbf{x}\in C_u$}%
}}}
\put(1726,-736){\makebox(0,0)[lb]{\smash{\SetFigFont{12}{14.4}{\familydefault}{\mddefault}{\updefault}{\color[rgb]{0,0,0}Encoder}%
}}}
\put(3076,-2161){\makebox(0,0)[lb]{\smash{\SetFigFont{12}{14.4}{\familydefault}{\mddefault}{\updefault}{\color[rgb]{0,0,0}Wire Tap}%
}}}
\put(3076,-2386){\makebox(0,0)[lb]{\smash{\SetFigFont{12}{14.4}{\familydefault}{\mddefault}{\updefault}{\color[rgb]{0,0,0}Channel}%
}}}
\put(4426,-736){\makebox(0,0)[lb]{\smash{\SetFigFont{12}{14.4}{\familydefault}{\mddefault}{\updefault}{\color[rgb]{0,0,0}Main}%
}}}
\put(4426,-961){\makebox(0,0)[lb]{\smash{\SetFigFont{12}{14.4}{\familydefault}{\mddefault}{\updefault}{\color[rgb]{0,0,0}Channel}%
}}}
\put(2851,-736){\makebox(0,0)[lb]{\smash{\SetFigFont{12}{14.4}{\familydefault}{\mddefault}{\updefault}{\color[rgb]{0,0,0}$\mathbf{x}$}%
}}}
\put(5551,-736){\makebox(0,0)[lb]{\smash{\SetFigFont{12}{14.4}{\familydefault}{\mddefault}{\updefault}{\color[rgb]{0,0,0}$\mathbf{y}$}%
}}}
\put(3526,-2911){\makebox(0,0)[lb]{\smash{\SetFigFont{12}{14.4}{\familydefault}{\mddefault}{\updefault}{\color[rgb]{0,0,0}$\mathbf{z}$}%
}}}
\put(6601,-886){\makebox(0,0)[lb]{\smash{\SetFigFont{12}{14.4}{\familydefault}{\mddefault}{\updefault}{\color[rgb]{0,0,0}Decoder}%
}}}
\put(826,-961){\makebox(0,0)[lb]{\smash{\SetFigFont{12}{14.4}{\familydefault}{\mddefault}{\updefault}{\color[rgb]{0,0,0}Alice}%
}}}
\put(7876,-961){\makebox(0,0)[lb]{\smash{\SetFigFont{12}{14.4}{\familydefault}{\mddefault}{\updefault}{\color[rgb]{0,0,0}Bob}%
}}}
\put(3076,-3061){\makebox(0,0)[lb]{\smash{\SetFigFont{12}{14.4}{\familydefault}{\mddefault}{\updefault}{\color[rgb]{0,0,0}Eve}%
}}}
\put(7726,-736){\makebox(0,0)[lb]{\smash{\SetFigFont{12}{14.4}{\familydefault}{\mddefault}{\updefault}{\color[rgb]{0,0,0}$\hat{\mathbf{u}}$}%
}}}
\put(976,-736){\makebox(0,0)[lb]{\smash{\SetFigFont{12}{14.4}{\familydefault}{\mddefault}{\updefault}{\color[rgb]{0,0,0}$\mathbf{u}$}%
}}}
\end{picture}

%% file: ewt.pstex_t
\begin{picture}(0,0)%
\includegraphics{ewt}%
\end{picture}%
\setlength{\unitlength}{3947sp}%
\begingroup\makeatletter\ifx\SetFigFont\undefined%
\gdef\SetFigFont#1#2#3#4#5{%
  \reset@font\fontsize{#1}{#2pt}%
  \fontfamily{#3}\fontseries{#4}\fontshape{#5}%
  \selectfont}%
\fi\endgroup%
\begin{picture}(3999,2067)(589,-1873)
\put(1201,-61){\makebox(0,0)[lb]{\smash{\SetFigFont{12}{14.4}{\familydefault}{\mddefault}{\updefault}{\color[rgb]{0,0,0}Alice}%
}}}
\put(2026, 14){\makebox(0,0)[lb]{\smash{\SetFigFont{12}{14.4}{\familydefault}{\mddefault}{\updefault}{\color[rgb]{0,0,0}$\mathbf{X}$}%
}}}
\put(751, 14){\makebox(0,0)[lb]{\smash{\SetFigFont{12}{14.4}{\familydefault}{\mddefault}{\updefault}{\color[rgb]{0,0,0}$\mathbf{S}$}%
}}}
\put(3976,-136){\makebox(0,0)[lb]{\smash{\SetFigFont{12}{14.4}{\familydefault}{\mddefault}{\updefault}{\color[rgb]{0,0,0}Bob}%
}}}
\put(3451, 14){\makebox(0,0)[lb]{\smash{\SetFigFont{12}{14.4}{\familydefault}{\mddefault}{\updefault}{\color[rgb]{0,0,0}$\mathbf{Y}$}%
}}}
\put(2701,-1711){\makebox(0,0)[lb]{\smash{\SetFigFont{12}{14.4}{\familydefault}{\mddefault}{\updefault}{\color[rgb]{0,0,0}Eve}%
}}}
\put(2926,-1186){\makebox(0,0)[lb]{\smash{\SetFigFont{12}{14.4}{\familydefault}{\mddefault}{\updefault}{\color[rgb]{0,0,0}$\mathbf{Z}$}%
}}}
\put(2371,-736){\makebox(0,0)[lb]{\smash{\SetFigFont{12}{14.4}{\familydefault}{\mddefault}{\updefault}{\color[rgb]{0,0,0}BEC($1-\epsilon$)}%
}}}
\end{picture}

%% file: wtc2.pstex_t
\begin{picture}(0,0)%
\includegraphics{wtc2}%
\end{picture}%
\setlength{\unitlength}{3947sp}%
\begingroup\makeatletter\ifx\SetFigFont\undefined%
\gdef\SetFigFont#1#2#3#4#5{%
  \reset@font\fontsize{#1}{#2pt}%
  \fontfamily{#3}\fontseries{#4}\fontshape{#5}%
  \selectfont}%
\fi\endgroup%
\begin{picture}(6924,2457)(589,-2881)
\put(4951,-736){\makebox(0,0)[lb]{\smash{\SetFigFont{12}{14.4}{\familydefault}{\mddefault}{\updefault}{\color[rgb]{0,0,0}$\mathbf{X}$}%
}}}
\put(6901,-736){\makebox(0,0)[lb]{\smash{\SetFigFont{12}{14.4}{\familydefault}{\mddefault}{\updefault}{\color[rgb]{0,0,0}$\mathbf{S}$}%
}}}
\put(6901,-1036){\makebox(0,0)[lb]{\smash{\SetFigFont{12}{14.4}{\familydefault}{\mddefault}{\updefault}{\color[rgb]{0,0,0}Bob}%
}}}
\put(4051,-2386){\makebox(0,0)[lb]{\smash{\SetFigFont{12}{14.4}{\familydefault}{\mddefault}{\updefault}{\color[rgb]{0,0,0}$\mathbf{Z}$}%
}}}
\put(3751,-2836){\makebox(0,0)[lb]{\smash{\SetFigFont{12}{14.4}{\familydefault}{\mddefault}{\updefault}{\color[rgb]{0,0,0}Eavesdropper}%
}}}
\put(2851,-736){\makebox(0,0)[lb]{\smash{\SetFigFont{12}{14.4}{\familydefault}{\mddefault}{\updefault}{\color[rgb]{0,0,0}$\mathbf{X}$}%
}}}
\put(901,-736){\makebox(0,0)[lb]{\smash{\SetFigFont{12}{14.4}{\familydefault}{\mddefault}{\updefault}{\color[rgb]{0,0,0}$\mathbf{S}$}%
}}}
\put(901,-1036){\makebox(0,0)[lb]{\smash{\SetFigFont{12}{14.4}{\familydefault}{\mddefault}{\updefault}{\color[rgb]{0,0,0}Alice}%
}}}
\put(5251,-1411){\makebox(0,0)[lb]{\smash{\SetFigFont{12}{14.4}{\familydefault}{\mddefault}{\updefault}{\color[rgb]{0,0,0}$H$: parity-check matrix of $\mathbf{C}$}%
}}}
\put(1771,-856){\makebox(0,0)[lb]{\smash{\SetFigFont{12}{14.4}{\familydefault}{\mddefault}{\updefault}{\color[rgb]{0,0,0}Encoder}%
}}}
\put(5611,-826){\makebox(0,0)[lb]{\smash{\SetFigFont{12}{14.4}{\familydefault}{\mddefault}{\updefault}{\color[rgb]{0,0,0}$\mathbf{S}=H\mathbf{X}^T$}%
}}}
\put(3526,-1561){\makebox(0,0)[lb]{\smash{\SetFigFont{12}{14.4}{\familydefault}{\mddefault}{\updefault}{\color[rgb]{0,0,0}BEC($1-\epsilon$)}%
}}}
\put(1426,-1606){\makebox(0,0)[lb]{\smash{\SetFigFont{12}{14.4}{\familydefault}{\mddefault}{\updefault}{\color[rgb]{0,0,0}seclected from}%
}}}
\put(1426,-1411){\makebox(0,0)[lb]{\smash{\SetFigFont{12}{14.4}{\familydefault}{\mddefault}{\updefault}{\color[rgb]{0,0,0}$\mathbf{X}$: Randomly }%
}}}
\put(1426,-1801){\makebox(0,0)[lb]{\smash{\SetFigFont{12}{14.4}{\familydefault}{\mddefault}{\updefault}{\color[rgb]{0,0,0}coset of a code $\mathbf{C}$}%
}}}
\put(1426,-2011){\makebox(0,0)[lb]{\smash{\SetFigFont{12}{14.4}{\familydefault}{\mddefault}{\updefault}{\color[rgb]{0,0,0} with syndrome $\mathbf{S}$}%
}}}
\end{picture}

%% file: initfig.pstex_t
\begin{picture}(0,0)%
\includegraphics{initfig}%
\end{picture}%
\setlength{\unitlength}{3947sp}%
\begingroup\makeatletter\ifx\SetFigFont\undefined%
\gdef\SetFigFont#1#2#3#4#5{%
  \reset@font\fontsize{#1}{#2pt}%
  \fontfamily{#3}\fontseries{#4}\fontshape{#5}%
  \selectfont}%
\fi\endgroup%
\begin{picture}(8325,4644)(1564,-5173)
\put(4651,-4261){\makebox(0,0)[lb]{\smash{{\SetFigFont{17}{20.4}{\rmdefault}{\mddefault}{\updefault}{\color[rgb]{0,0,0}$G^{*}$}%
}}}}
\put(4651,-2161){\makebox(0,0)[lb]{\smash{{\SetFigFont{17}{20.4}{\rmdefault}{\mddefault}{\updefault}{\color[rgb]{0,0,0}$G$}%
}}}}
\put(1801,-3586){\makebox(0,0)[lb]{\smash{{\SetFigFont{12}{14.4}{\rmdefault}{\mddefault}{\updefault}{\color[rgb]{0,0,0}$s_2$}%
}}}}
\put(1801,-2986){\makebox(0,0)[lb]{\smash{{\SetFigFont{12}{14.4}{\rmdefault}{\mddefault}{\updefault}{\color[rgb]{0,0,0}$v_{n(1-r)}$}%
}}}}
\put(1801,-1261){\makebox(0,0)[lb]{\smash{{\SetFigFont{12}{14.4}{\rmdefault}{\mddefault}{\updefault}{\color[rgb]{0,0,0}$v_1$}%
}}}}
\put(1801,-1561){\makebox(0,0)[lb]{\smash{{\SetFigFont{12}{14.4}{\rmdefault}{\mddefault}{\updefault}{\color[rgb]{0,0,0}$v_2$}%
}}}}
\put(1801,-3286){\makebox(0,0)[lb]{\smash{{\SetFigFont{12}{14.4}{\rmdefault}{\mddefault}{\updefault}{\color[rgb]{0,0,0}$s_1$}%
}}}}
\put(4876,-661){\makebox(0,0)[lb]{\smash{{\SetFigFont{12}{14.4}{\rmdefault}{\mddefault}{\updefault}{\color[rgb]{0,0,0}$n$}%
}}}}
\put(7951,-2236){\makebox(0,0)[lb]{\smash{{\SetFigFont{12}{14.4}{\rmdefault}{\mddefault}{\updefault}{\color[rgb]{0,0,0}$n(1-r)$}%
}}}}
\put(2176,-811){\makebox(0,0)[lb]{\smash{{\SetFigFont{14}{16.8}{\rmdefault}{\mddefault}{\updefault}{\color[rgb]{0,0,0}T}%
}}}}
\put(9751,-811){\makebox(0,0)[lb]{\smash{{\SetFigFont{14}{16.8}{\rmdefault}{\mddefault}{\updefault}{\color[rgb]{0,0,0}T}%
}}}}
\put(8326,-2986){\makebox(0,0)[lb]{\smash{{\SetFigFont{14}{16.8}{\rmdefault}{\mddefault}{\updefault}{\color[rgb]{0,0,0}=}%
}}}}
\put(9226,-1261){\makebox(0,0)[lb]{\smash{{\SetFigFont{12}{14.4}{\rmdefault}{\mddefault}{\updefault}{\color[rgb]{0,0,0}$x_1$}%
}}}}
\put(9226,-1561){\makebox(0,0)[lb]{\smash{{\SetFigFont{12}{14.4}{\rmdefault}{\mddefault}{\updefault}{\color[rgb]{0,0,0}$x_2$}%
}}}}
\put(9226,-4936){\makebox(0,0)[lb]{\smash{{\SetFigFont{12}{14.4}{\rmdefault}{\mddefault}{\updefault}{\color[rgb]{0,0,0}$x_n$}%
}}}}
\put(1801,-5011){\makebox(0,0)[lb]{\smash{{\SetFigFont{12}{14.4}{\rmdefault}{\mddefault}{\updefault}{\color[rgb]{0,0,0}$s_{nr}$}%
}}}}
\put(7951,-4261){\makebox(0,0)[lb]{\smash{{\SetFigFont{12}{14.4}{\rmdefault}{\mddefault}{\updefault}{\color[rgb]{0,0,0}$nr$}%
}}}}
\end{picture}%

%% file: hstar.pstex_t
\begin{picture}(0,0)%
\includegraphics{hstar}%
\end{picture}%
\setlength{\unitlength}{3947sp}%
\begingroup\makeatletter\ifx\SetFigFont\undefined%
\gdef\SetFigFont#1#2#3#4#5{%
  \reset@font\fontsize{#1}{#2pt}%
  \fontfamily{#3}\fontseries{#4}\fontshape{#5}%
  \selectfont}%
\fi\endgroup%
\begin{picture}(8325,4929)(1564,-5173)
\put(1801,-4336){\makebox(0,0)[lb]{\smash{{\SetFigFont{12}{14.4}{\rmdefault}{\mddefault}{\updefault}{\color[rgb]{0,0,0}$s_{n(r-t)}$}%
}}}}
\put(4651,-2161){\makebox(0,0)[lb]{\smash{{\SetFigFont{17}{20.4}{\rmdefault}{\mddefault}{\updefault}{\color[rgb]{0,0,0}$G$}%
}}}}
\put(1801,-3586){\makebox(0,0)[lb]{\smash{{\SetFigFont{12}{14.4}{\rmdefault}{\mddefault}{\updefault}{\color[rgb]{0,0,0}$s_2$}%
}}}}
\put(1801,-4561){\makebox(0,0)[lb]{\smash{{\SetFigFont{12}{14.4}{\rmdefault}{\mddefault}{\updefault}{\color[rgb]{0,0,0}0}%
}}}}
\put(1801,-5011){\makebox(0,0)[lb]{\smash{{\SetFigFont{12}{14.4}{\rmdefault}{\mddefault}{\updefault}{\color[rgb]{0,0,0}0}%
}}}}
\put(1801,-2986){\makebox(0,0)[lb]{\smash{{\SetFigFont{12}{14.4}{\rmdefault}{\mddefault}{\updefault}{\color[rgb]{0,0,0}$v_{n(1-r)}$}%
}}}}
\put(1801,-1261){\makebox(0,0)[lb]{\smash{{\SetFigFont{12}{14.4}{\rmdefault}{\mddefault}{\updefault}{\color[rgb]{0,0,0}$v_1$}%
}}}}
\put(1801,-1561){\makebox(0,0)[lb]{\smash{{\SetFigFont{12}{14.4}{\rmdefault}{\mddefault}{\updefault}{\color[rgb]{0,0,0}$v_2$}%
}}}}
\put(1801,-3286){\makebox(0,0)[lb]{\smash{{\SetFigFont{12}{14.4}{\rmdefault}{\mddefault}{\updefault}{\color[rgb]{0,0,0}$s_1$}%
}}}}
\put(4876,-661){\makebox(0,0)[lb]{\smash{{\SetFigFont{12}{14.4}{\rmdefault}{\mddefault}{\updefault}{\color[rgb]{0,0,0}$n$}%
}}}}
\put(7951,-2236){\makebox(0,0)[lb]{\smash{{\SetFigFont{12}{14.4}{\rmdefault}{\mddefault}{\updefault}{\color[rgb]{0,0,0}$n(1-r)$}%
}}}}
\put(7951,-3811){\makebox(0,0)[lb]{\smash{{\SetFigFont{12}{14.4}{\rmdefault}{\mddefault}{\updefault}{\color[rgb]{0,0,0}$n(r-t)$}%
}}}}
\put(7951,-4936){\makebox(0,0)[lb]{\smash{{\SetFigFont{12}{14.4}{\rmdefault}{\mddefault}{\updefault}{\color[rgb]{0,0,0}$nt$}%
}}}}
\put(4651,-3886){\makebox(0,0)[lb]{\smash{{\SetFigFont{17}{20.4}{\rmdefault}{\mddefault}{\updefault}{\color[rgb]{0,0,0}$G_1$}%
}}}}
\put(4651,-4936){\makebox(0,0)[lb]{\smash{{\SetFigFont{17}{20.4}{\rmdefault}{\mddefault}{\updefault}{\color[rgb]{0,0,0}$P$}%
}}}}
\put(8626,-436){\makebox(0,0)[lb]{\smash{{\SetFigFont{17}{20.4}{\rmdefault}{\mddefault}{\updefault}{\color[rgb]{0,0,0}$H*$}%
}}}}
\put(2176,-811){\makebox(0,0)[lb]{\smash{{\SetFigFont{14}{16.8}{\rmdefault}{\mddefault}{\updefault}{\color[rgb]{0,0,0}T}%
}}}}
\put(9751,-811){\makebox(0,0)[lb]{\smash{{\SetFigFont{14}{16.8}{\rmdefault}{\mddefault}{\updefault}{\color[rgb]{0,0,0}T}%
}}}}
\put(8326,-2986){\makebox(0,0)[lb]{\smash{{\SetFigFont{14}{16.8}{\rmdefault}{\mddefault}{\updefault}{\color[rgb]{0,0,0}=}%
}}}}
\put(9226,-1261){\makebox(0,0)[lb]{\smash{{\SetFigFont{12}{14.4}{\rmdefault}{\mddefault}{\updefault}{\color[rgb]{0,0,0}$x_1$}%
}}}}
\put(9226,-1561){\makebox(0,0)[lb]{\smash{{\SetFigFont{12}{14.4}{\rmdefault}{\mddefault}{\updefault}{\color[rgb]{0,0,0}$x_2$}%
}}}}
\put(9226,-4936){\makebox(0,0)[lb]{\smash{{\SetFigFont{12}{14.4}{\rmdefault}{\mddefault}{\updefault}{\color[rgb]{0,0,0}$x_n$}%
}}}}
\end{picture}%

%% file: hbeta.pstex_t
\begin{picture}(0,0)%
\includegraphics{hbeta}%
\end{picture}%
\setlength{\unitlength}{3947sp}%
\begingroup\makeatletter\ifx\SetFigFont\undefined%
\gdef\SetFigFont#1#2#3#4#5{%
  \reset@font\fontsize{#1}{#2pt}%
  \fontfamily{#3}\fontseries{#4}\fontshape{#5}%
  \selectfont}%
\fi\endgroup%
\begin{picture}(8325,5199)(1564,-5173)
\put(4201,-736){\makebox(0,0)[lb]{\smash{{\SetFigFont{12}{14.4}{\rmdefault}{\mddefault}{\updefault}{\color[rgb]{0,0,0}$n(1-t)$}%
}}}}
\put(4651,-2161){\makebox(0,0)[lb]{\smash{{\SetFigFont{17}{20.4}{\rmdefault}{\mddefault}{\updefault}{\color[rgb]{0,0,0}$G$}%
}}}}
\put(1801,-3586){\makebox(0,0)[lb]{\smash{{\SetFigFont{12}{14.4}{\rmdefault}{\mddefault}{\updefault}{\color[rgb]{0,0,0}$s_2$}%
}}}}
\put(1801,-4561){\makebox(0,0)[lb]{\smash{{\SetFigFont{12}{14.4}{\rmdefault}{\mddefault}{\updefault}{\color[rgb]{0,0,0}0}%
}}}}
\put(1801,-5011){\makebox(0,0)[lb]{\smash{{\SetFigFont{12}{14.4}{\rmdefault}{\mddefault}{\updefault}{\color[rgb]{0,0,0}0}%
}}}}
\put(1801,-2986){\makebox(0,0)[lb]{\smash{{\SetFigFont{12}{14.4}{\rmdefault}{\mddefault}{\updefault}{\color[rgb]{0,0,0}$v_{n(1-r)}$}%
}}}}
\put(1801,-1261){\makebox(0,0)[lb]{\smash{{\SetFigFont{12}{14.4}{\rmdefault}{\mddefault}{\updefault}{\color[rgb]{0,0,0}$v_1$}%
}}}}
\put(1801,-1561){\makebox(0,0)[lb]{\smash{{\SetFigFont{12}{14.4}{\rmdefault}{\mddefault}{\updefault}{\color[rgb]{0,0,0}$v_2$}%
}}}}
\put(1801,-3286){\makebox(0,0)[lb]{\smash{{\SetFigFont{12}{14.4}{\rmdefault}{\mddefault}{\updefault}{\color[rgb]{0,0,0}$s_1$}%
}}}}
\put(7951,-2236){\makebox(0,0)[lb]{\smash{{\SetFigFont{12}{14.4}{\rmdefault}{\mddefault}{\updefault}{\color[rgb]{0,0,0}$n(1-r)$}%
}}}}
\put(7951,-3811){\makebox(0,0)[lb]{\smash{{\SetFigFont{12}{14.4}{\rmdefault}{\mddefault}{\updefault}{\color[rgb]{0,0,0}$n(r-t)$}%
}}}}
\put(7951,-4936){\makebox(0,0)[lb]{\smash{{\SetFigFont{12}{14.4}{\rmdefault}{\mddefault}{\updefault}{\color[rgb]{0,0,0}$nt$}%
}}}}
\put(4651,-3886){\makebox(0,0)[lb]{\smash{{\SetFigFont{17}{20.4}{\rmdefault}{\mddefault}{\updefault}{\color[rgb]{0,0,0}$G_1$}%
}}}}
\put(4651,-4936){\makebox(0,0)[lb]{\smash{{\SetFigFont{17}{20.4}{\rmdefault}{\mddefault}{\updefault}{\color[rgb]{0,0,0}$P$}%
}}}}
\put(2176,-811){\makebox(0,0)[lb]{\smash{{\SetFigFont{14}{16.8}{\rmdefault}{\mddefault}{\updefault}{\color[rgb]{0,0,0}T}%
}}}}
\put(9751,-811){\makebox(0,0)[lb]{\smash{{\SetFigFont{14}{16.8}{\rmdefault}{\mddefault}{\updefault}{\color[rgb]{0,0,0}T}%
}}}}
\put(8326,-2986){\makebox(0,0)[lb]{\smash{{\SetFigFont{14}{16.8}{\rmdefault}{\mddefault}{\updefault}{\color[rgb]{0,0,0}=}%
}}}}
\put(9226,-1261){\makebox(0,0)[lb]{\smash{{\SetFigFont{12}{14.4}{\rmdefault}{\mddefault}{\updefault}{\color[rgb]{0,0,0}$x_1$}%
}}}}
\put(9226,-1561){\makebox(0,0)[lb]{\smash{{\SetFigFont{12}{14.4}{\rmdefault}{\mddefault}{\updefault}{\color[rgb]{0,0,0}$x_2$}%
}}}}
\put(9226,-4936){\makebox(0,0)[lb]{\smash{{\SetFigFont{12}{14.4}{\rmdefault}{\mddefault}{\updefault}{\color[rgb]{0,0,0}$x_n$}%
}}}}
\put(1801,-4336){\makebox(0,0)[lb]{\smash{{\SetFigFont{12}{14.4}{\rmdefault}{\mddefault}{\updefault}{\color[rgb]{0,0,0}$s_{n(r-t)}$}%
}}}}
\put(2926,-361){\makebox(0,0)[lb]{\smash{{\SetFigFont{12}{14.4}{\rmdefault}{\mddefault}{\updefault}{\color[rgb]{0,0,0}$n\beta$}%
}}}}
\put(6676,-286){\makebox(0,0)[lb]{\smash{{\SetFigFont{17}{20.4}{\rmdefault}{\mddefault}{\updefault}{\color[rgb]{0,0,0}$H_1$}%
}}}}
\put(2626,-2311){\makebox(0,0)[lb]{\smash{{\SetFigFont{12}{14.4}{\rmdefault}{\mddefault}{\updefault}{\color[rgb]{0,0,0}1}%
}}}}
\put(2926,-2011){\makebox(0,0)[lb]{\smash{{\SetFigFont{12}{14.4}{\rmdefault}{\mddefault}{\updefault}{\color[rgb]{0,0,0}1}%
}}}}
\put(2776,-1486){\makebox(0,0)[lb]{\smash{{\SetFigFont{12}{14.4}{\rmdefault}{\mddefault}{\updefault}{\color[rgb]{0,0,0}0}%
}}}}
\put(3226,-1786){\makebox(0,0)[lb]{\smash{{\SetFigFont{12}{14.4}{\rmdefault}{\mddefault}{\updefault}{\color[rgb]{0,0,0}1}%
}}}}
\put(3526,-1486){\makebox(0,0)[lb]{\smash{{\SetFigFont{12}{14.4}{\rmdefault}{\mddefault}{\updefault}{\color[rgb]{0,0,0}1}%
}}}}
\end{picture}%

%% file: hmat1.pstex_t
\begin{picture}(0,0)%
\includegraphics{hmat1}%
\end{picture}%
\setlength{\unitlength}{3947sp}%
\begingroup\makeatletter\ifx\SetFigFont\undefined%
\gdef\SetFigFont#1#2#3#4#5{%
  \reset@font\fontsize{#1}{#2pt}%
  \fontfamily{#3}\fontseries{#4}\fontshape{#5}%
  \selectfont}%
\fi\endgroup%
\begin{picture}(7200,3744)(2551,-4873)
\put(7201,-1861){\makebox(0,0)[lb]{\smash{{\SetFigFont{12}{14.4}{\rmdefault}{\mddefault}{\updefault}{\color[rgb]{0,0,0}$s_{n(r-t)}$}%
}}}}
\put(4201,-2536){\makebox(0,0)[lb]{\smash{{\SetFigFont{17}{20.4}{\rmdefault}{\mddefault}{\updefault}{\color[rgb]{0,0,0}B}%
}}}}
\put(5776,-2536){\makebox(0,0)[lb]{\smash{{\SetFigFont{17}{20.4}{\rmdefault}{\mddefault}{\updefault}{\color[rgb]{0,0,0}T}%
}}}}
\put(4276,-4111){\makebox(0,0)[lb]{\smash{{\SetFigFont{17}{20.4}{\rmdefault}{\mddefault}{\updefault}{\color[rgb]{0,0,0}D}%
}}}}
\put(5776,-4111){\makebox(0,0)[lb]{\smash{{\SetFigFont{17}{20.4}{\rmdefault}{\mddefault}{\updefault}{\color[rgb]{0,0,0}E}%
}}}}
\put(5551,-1261){\makebox(0,0)[lb]{\smash{{\SetFigFont{12}{14.4}{\rmdefault}{\mddefault}{\updefault}{\color[rgb]{0,0,0}$n\beta$}%
}}}}
\put(3751,-1336){\makebox(0,0)[lb]{\smash{{\SetFigFont{12}{14.4}{\rmdefault}{\mddefault}{\updefault}{\color[rgb]{0,0,0}$n(1-t-\beta)$}%
}}}}
\put(7201,-3286){\makebox(0,0)[lb]{\smash{{\SetFigFont{12}{14.4}{\rmdefault}{\mddefault}{\updefault}{\color[rgb]{0,0,0}$s_1$}%
}}}}
\put(7201,-3586){\makebox(0,0)[lb]{\smash{{\SetFigFont{12}{14.4}{\rmdefault}{\mddefault}{\updefault}{\color[rgb]{0,0,0}$v_{n(1-r)}$}%
}}}}
\put(7201,-4786){\makebox(0,0)[lb]{\smash{{\SetFigFont{12}{14.4}{\rmdefault}{\mddefault}{\updefault}{\color[rgb]{0,0,0}$v_1$}%
}}}}
\put(8551,-1786){\makebox(0,0)[lb]{\smash{{\SetFigFont{12}{14.4}{\rmdefault}{\mddefault}{\updefault}{\color[rgb]{0,0,0}$x_1$}%
}}}}
\put(8551,-2161){\makebox(0,0)[lb]{\smash{{\SetFigFont{12}{14.4}{\rmdefault}{\mddefault}{\updefault}{\color[rgb]{0,0,0}$x_2$}%
}}}}
\put(8551,-4336){\makebox(0,0)[lb]{\smash{{\SetFigFont{12}{14.4}{\rmdefault}{\mddefault}{\updefault}{\color[rgb]{0,0,0}$x_{n(1-t)}$}%
}}}}
\put(5926,-2011){\makebox(0,0)[lb]{\smash{{\SetFigFont{17}{20.4}{\rmdefault}{\mddefault}{\updefault}{\color[rgb]{0,0,0}$0$}%
}}}}
\put(7876,-2836){\makebox(0,0)[lb]{\smash{{\SetFigFont{14}{16.8}{\rmdefault}{\mddefault}{\updefault}{\color[rgb]{0,0,0}=}%
}}}}
\put(9751,-3286){\makebox(0,0)[lb]{\smash{{\SetFigFont{12}{14.4}{\rmdefault}{\mddefault}{\updefault}{\color[rgb]{0,0,0}$\mathbf{X^*}$ vector}%
}}}}
\put(2551,-4036){\makebox(0,0)[lb]{\smash{{\SetFigFont{12}{14.4}{\rmdefault}{\mddefault}{\updefault}{\color[rgb]{0,0,0}$n(1-t-\beta)$}%
}}}}
\put(3001,-2536){\makebox(0,0)[lb]{\smash{{\SetFigFont{12}{14.4}{\rmdefault}{\mddefault}{\updefault}{\color[rgb]{0,0,0}$n\beta$}%
}}}}
\end{picture}%

%% file: qmat.pstex_t
\begin{picture}(0,0)%
\includegraphics{qmat}%
\end{picture}%
\setlength{\unitlength}{3947sp}%
\begingroup\makeatletter\ifx\SetFigFont\undefined%
\gdef\SetFigFont#1#2#3#4#5{%
  \reset@font\fontsize{#1}{#2pt}%
  \fontfamily{#3}\fontseries{#4}\fontshape{#5}%
  \selectfont}%
\fi\endgroup%
\begin{picture}(3087,3018)(3589,-4198)
\put(4351,-3811){\makebox(0,0)[lb]{\smash{{\SetFigFont{17}{20.4}{\rmdefault}{\mddefault}{\updefault}{\color[rgb]{0,0,0}$ET^{-1}$}%
}}}}
\put(4351,-1336){\makebox(0,0)[lb]{\smash{{\SetFigFont{12}{14.4}{\rmdefault}{\mddefault}{\updefault}{\color[rgb]{0,0,0}$n\beta$}%
}}}}
\put(6676,-2461){\makebox(0,0)[lb]{\smash{{\SetFigFont{12}{14.4}{\rmdefault}{\mddefault}{\updefault}{\color[rgb]{0,0,0}$n\beta$}%
}}}}
\put(5401,-1336){\makebox(0,0)[lb]{\smash{{\SetFigFont{12}{14.4}{\rmdefault}{\mddefault}{\updefault}{\color[rgb]{0,0,0}$n(1-t-\beta)$}%
}}}}
\put(6676,-3736){\makebox(0,0)[lb]{\smash{{\SetFigFont{12}{14.4}{\rmdefault}{\mddefault}{\updefault}{\color[rgb]{0,0,0}$n(1-t-\beta)$}%
}}}}
\put(4351,-2386){\makebox(0,0)[lb]{\smash{{\SetFigFont{17}{20.4}{\rmdefault}{\mddefault}{\updefault}{\color[rgb]{0,0,0}$I$}%
}}}}
\put(5851,-3811){\makebox(0,0)[lb]{\smash{{\SetFigFont{17}{20.4}{\rmdefault}{\mddefault}{\updefault}{\color[rgb]{0,0,0}$I$}%
}}}}
\put(5701,-2386){\makebox(0,0)[lb]{\smash{{\SetFigFont{17}{20.4}{\rmdefault}{\mddefault}{\updefault}{\color[rgb]{0,0,0}$0$}%
}}}}
\end{picture}%

%% file: yvect.pstex_t
\begin{picture}(0,0)%
\includegraphics{yvect}%
\end{picture}%
\setlength{\unitlength}{3947sp}%
\begingroup\makeatletter\ifx\SetFigFont\undefined%
\gdef\SetFigFont#1#2#3#4#5{%
  \reset@font\fontsize{#1}{#2pt}%
  \fontfamily{#3}\fontseries{#4}\fontshape{#5}%
  \selectfont}%
\fi\endgroup%
\begin{picture}(7125,3963)(2626,-5092)
\put(7126,-1861){\makebox(0,0)[lb]{\smash{{\SetFigFont{12}{14.4}{\rmdefault}{\mddefault}{\updefault}{\color[rgb]{0,0,0}$s_{n(r-t)}$}%
}}}}
\put(4201,-2536){\makebox(0,0)[lb]{\smash{{\SetFigFont{17}{20.4}{\rmdefault}{\mddefault}{\updefault}{\color[rgb]{0,0,0}B}%
}}}}
\put(5776,-2536){\makebox(0,0)[lb]{\smash{{\SetFigFont{17}{20.4}{\rmdefault}{\mddefault}{\updefault}{\color[rgb]{0,0,0}T}%
}}}}
\put(5551,-1261){\makebox(0,0)[lb]{\smash{{\SetFigFont{12}{14.4}{\rmdefault}{\mddefault}{\updefault}{\color[rgb]{0,0,0}$n\beta$}%
}}}}
\put(3751,-1336){\makebox(0,0)[lb]{\smash{{\SetFigFont{12}{14.4}{\rmdefault}{\mddefault}{\updefault}{\color[rgb]{0,0,0}$n(1-t-\beta)$}%
}}}}
\put(7201,-3286){\makebox(0,0)[lb]{\smash{{\SetFigFont{12}{14.4}{\rmdefault}{\mddefault}{\updefault}{\color[rgb]{0,0,0}$s_1$}%
}}}}
\put(7201,-3586){\makebox(0,0)[lb]{\smash{{\SetFigFont{12}{14.4}{\rmdefault}{\mddefault}{\updefault}{\color[rgb]{0,0,0}$v_{n(1-r)}$}%
}}}}
\put(7201,-4786){\makebox(0,0)[lb]{\smash{{\SetFigFont{12}{14.4}{\rmdefault}{\mddefault}{\updefault}{\color[rgb]{0,0,0}$v_1$}%
}}}}
\put(7876,-2836){\makebox(0,0)[lb]{\smash{{\SetFigFont{14}{16.8}{\rmdefault}{\mddefault}{\updefault}{\color[rgb]{0,0,0}=}%
}}}}
\put(5776,-4111){\makebox(0,0)[lb]{\smash{{\SetFigFont{17}{20.4}{\rmdefault}{\mddefault}{\updefault}{\color[rgb]{0,0,0}0}%
}}}}
\put(8551,-1786){\makebox(0,0)[lb]{\smash{{\SetFigFont{12}{14.4}{\rmdefault}{\mddefault}{\updefault}{\color[rgb]{0,0,0}$y_1$}%
}}}}
\put(8551,-2161){\makebox(0,0)[lb]{\smash{{\SetFigFont{12}{14.4}{\rmdefault}{\mddefault}{\updefault}{\color[rgb]{0,0,0}$y_2$}%
}}}}
\put(8551,-4336){\makebox(0,0)[lb]{\smash{{\SetFigFont{12}{14.4}{\rmdefault}{\mddefault}{\updefault}{\color[rgb]{0,0,0}$y_{n(1-t)}$}%
}}}}
\put(9751,-3286){\makebox(0,0)[lb]{\smash{{\SetFigFont{12}{14.4}{\rmdefault}{\mddefault}{\updefault}{\color[rgb]{0,0,0}$\mathbf{Y}$ vector}%
}}}}
\put(3001,-2611){\makebox(0,0)[lb]{\smash{{\SetFigFont{12}{14.4}{\rmdefault}{\mddefault}{\updefault}{\color[rgb]{0,0,0}$n\beta$}%
}}}}
\put(2626,-4111){\makebox(0,0)[lb]{\smash{{\SetFigFont{12}{14.4}{\rmdefault}{\mddefault}{\updefault}{\color[rgb]{0,0,0}$n(1-t-\beta)$}%
}}}}
\put(5926,-2086){\makebox(0,0)[lb]{\smash{{\SetFigFont{17}{20.4}{\rmdefault}{\mddefault}{\updefault}{\color[rgb]{0,0,0}$0$}%
}}}}
\put(3751,-5011){\makebox(0,0)[lb]{\smash{{\SetFigFont{17}{20.4}{\rmdefault}{\mddefault}{\updefault}{\color[rgb]{0,0,0}E$T^{-1}$B+D}%
}}}}
\end{picture}%

%% file: y12.pstex_t
\begin{picture}(0,0)%
\includegraphics{y12}%
\end{picture}%
\setlength{\unitlength}{3947sp}%
\begingroup\makeatletter\ifx\SetFigFont\undefined%
\gdef\SetFigFont#1#2#3#4#5{%
  \reset@font\fontsize{#1}{#2pt}%
  \fontfamily{#3}\fontseries{#4}\fontshape{#5}%
  \selectfont}%
\fi\endgroup%
\begin{picture}(7090,4864)(376,-5294)
\put(5701,-586){\makebox(0,0)[lb]{\smash{{\SetFigFont{12}{14.4}{\rmdefault}{\mddefault}{\updefault}{\color[rgb]{0,0,0}$\mathbf{X^*_1}$ vector}%
}}}}
\put(1801,-1261){\makebox(0,0)[lb]{\smash{{\SetFigFont{12}{14.4}{\rmdefault}{\mddefault}{\updefault}{\color[rgb]{0,0,0}$n\beta$}%
}}}}
\put(3151,-1261){\makebox(0,0)[lb]{\smash{{\SetFigFont{12}{14.4}{\rmdefault}{\mddefault}{\updefault}{\color[rgb]{0,0,0}$n(1-t-\beta)$}%
}}}}
\put(1951,-2386){\makebox(0,0)[lb]{\smash{{\SetFigFont{17}{20.4}{\rmdefault}{\mddefault}{\updefault}{\color[rgb]{0,0,0}$I$}%
}}}}
\put(1876,-3736){\makebox(0,0)[lb]{\smash{{\SetFigFont{17}{20.4}{\rmdefault}{\mddefault}{\updefault}{\color[rgb]{0,0,0}$ET^{-1}$}%
}}}}
\put(3451,-3736){\makebox(0,0)[lb]{\smash{{\SetFigFont{17}{20.4}{\rmdefault}{\mddefault}{\updefault}{\color[rgb]{0,0,0}$I$}%
}}}}
\put(3451,-2386){\makebox(0,0)[lb]{\smash{{\SetFigFont{17}{20.4}{\rmdefault}{\mddefault}{\updefault}{\color[rgb]{0,0,0}$0$}%
}}}}
\put(676,-2386){\makebox(0,0)[lb]{\smash{{\SetFigFont{12}{14.4}{\rmdefault}{\mddefault}{\updefault}{\color[rgb]{0,0,0}$n\beta$}%
}}}}
\put(376,-3811){\makebox(0,0)[lb]{\smash{{\SetFigFont{12}{14.4}{\rmdefault}{\mddefault}{\updefault}{\color[rgb]{0,0,0}$n(1-t-\beta)$}%
}}}}
\put(2026,-4636){\makebox(0,0)[lb]{\smash{{\SetFigFont{17}{20.4}{\rmdefault}{\mddefault}{\updefault}{\color[rgb]{0,0,0}Q matrix}%
}}}}
\put(4651,-1261){\makebox(0,0)[lb]{\smash{{\SetFigFont{12}{14.4}{\rmdefault}{\mddefault}{\updefault}{\color[rgb]{0,0,0}$x_1$}%
}}}}
\put(4651,-1561){\makebox(0,0)[lb]{\smash{{\SetFigFont{12}{14.4}{\rmdefault}{\mddefault}{\updefault}{\color[rgb]{0,0,0}$x_2$}%
}}}}
\put(4726,-4486){\makebox(0,0)[lb]{\smash{{\SetFigFont{12}{14.4}{\rmdefault}{\mddefault}{\updefault}{\color[rgb]{0,0,0}$x_{n(1-t)}$}%
}}}}
\put(6451,-1261){\makebox(0,0)[lb]{\smash{{\SetFigFont{12}{14.4}{\rmdefault}{\mddefault}{\updefault}{\color[rgb]{0,0,0}$y_1$}%
}}}}
\put(6451,-1561){\makebox(0,0)[lb]{\smash{{\SetFigFont{12}{14.4}{\rmdefault}{\mddefault}{\updefault}{\color[rgb]{0,0,0}$y_2$}%
}}}}
\put(6451,-4411){\makebox(0,0)[lb]{\smash{{\SetFigFont{12}{14.4}{\rmdefault}{\mddefault}{\updefault}{\color[rgb]{0,0,0}$y_{n(1-t)}$}%
}}}}
\put(7276,-1936){\makebox(0,0)[lb]{\smash{{\SetFigFont{12}{14.4}{\rmdefault}{\mddefault}{\updefault}{\color[rgb]{0,0,0}$n\beta$}%
}}}}
\put(7276,-3661){\makebox(0,0)[lb]{\smash{{\SetFigFont{12}{14.4}{\rmdefault}{\mddefault}{\updefault}{\color[rgb]{0,0,0}$n(1-t-\beta)$}%
}}}}
\put(5326,-5236){\makebox(0,0)[lb]{\smash{{\SetFigFont{12}{14.4}{\rmdefault}{\mddefault}{\updefault}{\color[rgb]{0,0,0}$\mathbf{X^*_2}$ vector}%
}}}}
\put(7051,-5011){\makebox(0,0)[lb]{\smash{{\SetFigFont{12}{14.4}{\rmdefault}{\mddefault}{\updefault}{\color[rgb]{0,0,0}$\mathbf{Y_2}$ vector}%
}}}}
\put(3451,-4936){\makebox(0,0)[lb]{\smash{{\SetFigFont{12}{14.4}{\rmdefault}{\mddefault}{\updefault}{\color[rgb]{0,0,0}$\mathbf{X^*}$ vector}%
}}}}
\put(7276,-736){\makebox(0,0)[lb]{\smash{{\SetFigFont{12}{14.4}{\rmdefault}{\mddefault}{\updefault}{\color[rgb]{0,0,0}$\mathbf{Y_1}$ vector}%
}}}}
\end{picture}%

%% file: u12.pstex_t
\begin{picture}(0,0)%
\includegraphics{u12}%
\end{picture}%
\setlength{\unitlength}{3947sp}%
\begingroup\makeatletter\ifx\SetFigFont\undefined%
\gdef\SetFigFont#1#2#3#4#5{%
  \reset@font\fontsize{#1}{#2pt}%
  \fontfamily{#3}\fontseries{#4}\fontshape{#5}%
  \selectfont}%
\fi\endgroup%
\begin{picture}(7050,4114)(2701,-5219)
\put(3751,-5086){\makebox(0,0)[lb]{\smash{{\SetFigFont{17}{20.4}{\rmdefault}{\mddefault}{\updefault}{\color[rgb]{0,0,0}E$T^{-1}$B+D}%
}}}}
\put(4201,-2536){\makebox(0,0)[lb]{\smash{{\SetFigFont{17}{20.4}{\rmdefault}{\mddefault}{\updefault}{\color[rgb]{0,0,0}B}%
}}}}
\put(5776,-2536){\makebox(0,0)[lb]{\smash{{\SetFigFont{17}{20.4}{\rmdefault}{\mddefault}{\updefault}{\color[rgb]{0,0,0}T}%
}}}}
\put(5551,-1261){\makebox(0,0)[lb]{\smash{{\SetFigFont{12}{14.4}{\rmdefault}{\mddefault}{\updefault}{\color[rgb]{0,0,0}$n\beta$}%
}}}}
\put(3751,-1336){\makebox(0,0)[lb]{\smash{{\SetFigFont{12}{14.4}{\rmdefault}{\mddefault}{\updefault}{\color[rgb]{0,0,0}$n(1-t-\beta)$}%
}}}}
\put(7201,-2311){\makebox(0,0)[lb]{\smash{{\SetFigFont{12}{14.4}{\rmdefault}{\mddefault}{\updefault}{\color[rgb]{0,0,0}$s_{n(r-t)}$}%
}}}}
\put(7201,-3286){\makebox(0,0)[lb]{\smash{{\SetFigFont{12}{14.4}{\rmdefault}{\mddefault}{\updefault}{\color[rgb]{0,0,0}$s_1$}%
}}}}
\put(7201,-3586){\makebox(0,0)[lb]{\smash{{\SetFigFont{12}{14.4}{\rmdefault}{\mddefault}{\updefault}{\color[rgb]{0,0,0}$v_{n(1-r)}$}%
}}}}
\put(7201,-4786){\makebox(0,0)[lb]{\smash{{\SetFigFont{12}{14.4}{\rmdefault}{\mddefault}{\updefault}{\color[rgb]{0,0,0}$v_1$}%
}}}}
\put(5926,-2011){\makebox(0,0)[lb]{\smash{{\SetFigFont{17}{20.4}{\rmdefault}{\mddefault}{\updefault}{\color[rgb]{0,0,0}$0$}%
}}}}
\put(7876,-2836){\makebox(0,0)[lb]{\smash{{\SetFigFont{14}{16.8}{\rmdefault}{\mddefault}{\updefault}{\color[rgb]{0,0,0}=}%
}}}}
\put(5776,-4111){\makebox(0,0)[lb]{\smash{{\SetFigFont{17}{20.4}{\rmdefault}{\mddefault}{\updefault}{\color[rgb]{0,0,0}0}%
}}}}
\put(8551,-1786){\makebox(0,0)[lb]{\smash{{\SetFigFont{12}{14.4}{\rmdefault}{\mddefault}{\updefault}{\color[rgb]{0,0,0}$y_1$}%
}}}}
\put(8551,-2161){\makebox(0,0)[lb]{\smash{{\SetFigFont{12}{14.4}{\rmdefault}{\mddefault}{\updefault}{\color[rgb]{0,0,0}$y_2$}%
}}}}
\put(8551,-4336){\makebox(0,0)[lb]{\smash{{\SetFigFont{12}{14.4}{\rmdefault}{\mddefault}{\updefault}{\color[rgb]{0,0,0}$y_{n(1-t)}$}%
}}}}
\put(8251,-1261){\makebox(0,0)[lb]{\smash{{\SetFigFont{12}{14.4}{\rmdefault}{\mddefault}{\updefault}{\color[rgb]{0,0,0}$\mathbf{U_1}$ vector}%
}}}}
\put(8026,-5161){\makebox(0,0)[lb]{\smash{{\SetFigFont{12}{14.4}{\rmdefault}{\mddefault}{\updefault}{\color[rgb]{0,0,0}$\mathbf{U_2}$ vector}%
}}}}
\put(9751,-3211){\makebox(0,0)[lb]{\smash{{\SetFigFont{12}{14.4}{\rmdefault}{\mddefault}{\updefault}{\color[rgb]{0,0,0}$\mathbf{Y}$ vector}%
}}}}
\put(3001,-2536){\makebox(0,0)[lb]{\smash{{\SetFigFont{12}{14.4}{\rmdefault}{\mddefault}{\updefault}{\color[rgb]{0,0,0}$n\beta$}%
}}}}
\put(2701,-4036){\makebox(0,0)[lb]{\smash{{\SetFigFont{12}{14.4}{\rmdefault}{\mddefault}{\updefault}{\color[rgb]{0,0,0}$n(1-t-\beta)$}%
}}}}
\end{picture}%

%% file: gen_bec.pstex_t
\begin{picture}(0,0)%
\includegraphics{gen_bec}%
\end{picture}%
\setlength{\unitlength}{3947sp}%
\begingroup\makeatletter\ifx\SetFigFont\undefined%
\gdef\SetFigFont#1#2#3#4#5{%
  \reset@font\fontsize{#1}{#2pt}%
  \fontfamily{#3}\fontseries{#4}\fontshape{#5}%
  \selectfont}%
\fi\endgroup%
\begin{picture}(4524,2067)(589,-1873)
\put(2026,-736){\makebox(0,0)[lb]{\smash{\SetFigFont{12}{14.4}{\familydefault}{\mddefault}{\updefault}{\color[rgb]{0,0,0}BEC($\epsilon_w$)}%
}}}
\put(1201,-61){\makebox(0,0)[lb]{\smash{\SetFigFont{12}{14.4}{\familydefault}{\mddefault}{\updefault}{\color[rgb]{0,0,0}Alice}%
}}}
\put(2026, 14){\makebox(0,0)[lb]{\smash{\SetFigFont{12}{14.4}{\familydefault}{\mddefault}{\updefault}{\color[rgb]{0,0,0}$\mathbf{X}$}%
}}}
\put(2251,-1711){\makebox(0,0)[lb]{\smash{\SetFigFont{12}{14.4}{\familydefault}{\mddefault}{\updefault}{\color[rgb]{0,0,0}Eve}%
}}}
\put(4501,-136){\makebox(0,0)[lb]{\smash{\SetFigFont{12}{14.4}{\familydefault}{\mddefault}{\updefault}{\color[rgb]{0,0,0}Bob}%
}}}
\put(2476,-1186){\makebox(0,0)[lb]{\smash{\SetFigFont{12}{14.4}{\familydefault}{\mddefault}{\updefault}{\color[rgb]{0,0,0}$\mathbf{Z}$}%
}}}
\put(751, 14){\makebox(0,0)[lb]{\smash{\SetFigFont{12}{14.4}{\familydefault}{\mddefault}{\updefault}{\color[rgb]{0,0,0}$\mathbf{S}$}%
}}}
\put(3976, 14){\makebox(0,0)[lb]{\smash{\SetFigFont{12}{14.4}{\familydefault}{\mddefault}{\updefault}{\color[rgb]{0,0,0}$\mathbf{Y}$}%
}}}
\put(3001,-136){\makebox(0,0)[lb]{\smash{\SetFigFont{12}{14.4}{\familydefault}{\mddefault}{\updefault}{\color[rgb]{0,0,0}BEC($\epsilon_m$)}%
}}}
\end{picture}

%% file: Hmat.pstex_t
\begin{picture}(0,0)%
\includegraphics{Hmat}%
\end{picture}%
\setlength{\unitlength}{3947sp}%
\begingroup\makeatletter\ifx\SetFigFont\undefined%
\gdef\SetFigFont#1#2#3#4#5{%
  \reset@font\fontsize{#1}{#2pt}%
  \fontfamily{#3}\fontseries{#4}\fontshape{#5}%
  \selectfont}%
\fi\endgroup%
\begin{picture}(4962,2424)(226,-2173)
\put(4951,-586){\makebox(0,0)[lb]{\smash{\SetFigFont{10}{12.0}{\familydefault}{\mddefault}{\updefault}{\color[rgb]{0,0,0}$\mathbf{0}$}%
}}}
\put(2176,-1786){\makebox(0,0)[lb]{\smash{\SetFigFont{12}{14.4}{\familydefault}{\mddefault}{\updefault}{\color[rgb]{0,0,0}$\overline{H}_2$}%
}}}
\put(2176,-586){\makebox(0,0)[lb]{\smash{\SetFigFont{12}{14.4}{\familydefault}{\mddefault}{\updefault}{\color[rgb]{0,0,0}$H_2$}%
}}}
\put(4951,-1786){\makebox(0,0)[lb]{\smash{\SetFigFont{10}{12.0}{\familydefault}{\mddefault}{\updefault}{\color[rgb]{0,0,0}$\mathbf{S}^T$}%
}}}
\put(3976,-961){\makebox(0,0)[lb]{\smash{\SetFigFont{10}{12.0}{\familydefault}{\mddefault}{\updefault}{\color[rgb]{0,0,0} $\mathbf{X}^T$}%
}}}
\put(226,-1711){\makebox(0,0)[lb]{\smash{\SetFigFont{10}{12.0}{\familydefault}{\mddefault}{\updefault}{\color[rgb]{0,0,0}$n(r_2-r_1)$}%
}}}
\put(226,-511){\makebox(0,0)[lb]{\smash{\SetFigFont{10}{12.0}{\familydefault}{\mddefault}{\updefault}{\color[rgb]{0,0,0}$n(1-r_2)$}%
}}}
\end{picture}

%% file: enc_space.pstex_t
\begin{picture}(0,0)%
\includegraphics{enc_space}%
\end{picture}%
\setlength{\unitlength}{3947sp}%
\begingroup\makeatletter\ifx\SetFigFont\undefined%
\gdef\SetFigFont#1#2#3#4#5{%
  \reset@font\fontsize{#1}{#2pt}%
  \fontfamily{#3}\fontseries{#4}\fontshape{#5}%
  \selectfont}%
\fi\endgroup%
\begin{picture}(4362,3106)(1051,-3335)
\put(1051,-3286){\makebox(0,0)[lb]{\smash{\SetFigFont{10}{12.0}{\familydefault}{\mddefault}{\updefault}{\color[rgb]{0,0,0}solution set of $\overline{H}_2\mathbf{X}^T=\mathbf{S}^T_1$}%
}}}
\put(4576,-1036){\makebox(0,0)[lb]{\smash{\SetFigFont{10}{12.0}{\familydefault}{\mddefault}{\updefault}{\color[rgb]{0,0,0}for $\mathbf{S}=\mathbf{S_2}$}%
}}}
\put(4276,-886){\makebox(0,0)[lb]{\smash{\SetFigFont{10}{12.0}{\familydefault}{\mddefault}{\updefault}{\color[rgb]{0,0,0}solution set of Figure \ref{fig:Hmat}}%
}}}
\put(2401,-361){\makebox(0,0)[lb]{\smash{\SetFigFont{10}{12.0}{\familydefault}{\mddefault}{\updefault}{\color[rgb]{0,0,0}solution set of $H_2\mathbf{X}^T=\mathbf{0}$}%
}}}
\put(3601,-2836){\makebox(0,0)[lb]{\smash{\SetFigFont{10}{12.0}{\familydefault}{\mddefault}{\updefault}{\color[rgb]{0,0,0}for $\mathbf{S}=\mathbf{S_1}$}%
}}}
\put(3376,-2686){\makebox(0,0)[lb]{\smash{\SetFigFont{10}{12.0}{\familydefault}{\mddefault}{\updefault}{\color[rgb]{0,0,0}solution set of Figure \ref{fig:Hmat}}%
}}}
\put(4951,-2161){\makebox(0,0)[lb]{\smash{\SetFigFont{10}{12.0}{\familydefault}{\mddefault}{\updefault}{\color[rgb]{0,0,0}solution set of $\overline{H}_2\mathbf{X}^T=\mathbf{S}^T_2$}%
}}}
\end{picture}

%% file: BSC_EF.pstex_t
\begin{picture}(0,0)%
\includegraphics{BSC_EF}%
\end{picture}%
\setlength{\unitlength}{3947sp}%
\begingroup\makeatletter\ifx\SetFigFont\undefined%
\gdef\SetFigFont#1#2#3#4#5{%
  \reset@font\fontsize{#1}{#2pt}%
  \fontfamily{#3}\fontseries{#4}\fontshape{#5}%
  \selectfont}%
\fi\endgroup%
\begin{picture}(4899,2274)(289,-2623)
\put(301,-886){\makebox(0,0)[lb]{\smash{\SetFigFont{12}{14.4}{\familydefault}{\mddefault}{\updefault}{\color[rgb]{0,0,0}Alice}%
}}}
\put(2010,-623){\makebox(0,0)[lb]{\smash{\SetFigFont{12}{14.4}{\familydefault}{\mddefault}{\updefault}{\color[rgb]{0,0,0}Encoder}%
}}}
\put(451,-586){\makebox(0,0)[lb]{\smash{\SetFigFont{12}{14.4}{\familydefault}{\mddefault}{\updefault}{\color[rgb]{0,0,0}$\mathbf{S}$}%
}}}
\put(1126,-1036){\makebox(0,0)[lb]{\smash{\SetFigFont{12}{14.4}{\familydefault}{\mddefault}{\updefault}{\color[rgb]{0,0,0}$\mathbf{V}$}%
}}}
\put(3076,-586){\makebox(0,0)[lb]{\smash{\SetFigFont{12}{14.4}{\familydefault}{\mddefault}{\updefault}{\color[rgb]{0,0,0}$\mathbf{X}$}%
}}}
\put(3601,-1711){\makebox(0,0)[lb]{\smash{\SetFigFont{12}{14.4}{\familydefault}{\mddefault}{\updefault}{\color[rgb]{0,0,0}BSC($p_w$)}%
}}}
\put(3601,-2536){\makebox(0,0)[lb]{\smash{\SetFigFont{12}{14.4}{\familydefault}{\mddefault}{\updefault}{\color[rgb]{0,0,0}Eve}%
}}}
\put(4051,-2536){\makebox(0,0)[lb]{\smash{\SetFigFont{12}{14.4}{\familydefault}{\mddefault}{\updefault}{\color[rgb]{0,0,0}$\mathbf{Z}$}%
}}}
\put(4876,-886){\makebox(0,0)[lb]{\smash{\SetFigFont{12}{14.4}{\familydefault}{\mddefault}{\updefault}{\color[rgb]{0,0,0}Bob}%
}}}
\put(4876,-586){\makebox(0,0)[lb]{\smash{\SetFigFont{12}{14.4}{\familydefault}{\mddefault}{\updefault}{\color[rgb]{0,0,0}$\mathbf{X}$}%
}}}
\put(646,-1651){\makebox(0,0)[lb]{\smash{\SetFigFont{12}{14.4}{\familydefault}{\mddefault}{\updefault}{\color[rgb]{0,0,0}Generator}%
}}}
\put(2243,-833){\makebox(0,0)[lb]{\smash{\SetFigFont{12}{14.4}{\familydefault}{\mddefault}{\updefault}{\color[rgb]{0,0,0}$\overline{G}$}%
}}}
\put(533,-1426){\makebox(0,0)[lb]{\smash{\SetFigFont{12}{14.4}{\familydefault}{\mddefault}{\updefault}{\color[rgb]{0,0,0}Random bit}%
}}}
\end{picture}